\documentclass[pre,aps,floats,twocolumn,superscriptaddress,floatfix]{revtex4}
\usepackage{graphicx,amssymb,amsmath,ifthen,rotating}
\usepackage[active]{srcltx}
\begin{document}
\title{Density profiles of a self-gravitating lattice gas in one, two, and three dimensions}  
\author{Benaoumeur Bakhti}
\affiliation{
  Fachbereich Physik,
  Universit\"at Osnabr\"uck,
  D-49076 Osnabr\"uck, Germany}
  \author{Divana Boukari}
\affiliation{
  Department of Physics,
  University of Rhode Island,
  Kingston RI 02881, USA}
    \author{Michael Karbach}
\affiliation{
  Fachbereich Physik,
  Bergische Universit\"at Wuppertal,
  D-42097 Wuppertal, Germany}
    \author{Philipp Maass}
\affiliation{
  Fachbereich Physik,
  Universit\"at Osnabr\"uck,
  D-49076 Osnabr\"uck, Germany}
\author{Gerhard M{\"{u}}ller}
\affiliation{
  Department of Physics,
  University of Rhode Island,
  Kingston RI 02881, USA}

\begin{abstract}
We consider a lattice gas in spaces of dimensionality $\mathcal{D}=1,2,3$.
The particles are subject to a hardcore exclusion interaction and an attractive pair interaction that satisfies Gauss' law as do Newtonian gravity in $\mathcal{D}=3$, a logarithmic potential in $\mathcal{D}=2$, and a distance-independent force in $\mathcal{D}=1$.
Under mild additional assumptions regarding symmetry and fluctuations we investigate equilibrium states of self-gravitating material clusters, in particular radial density profiles for closed and open systems.
We present exact analytic results in several instances and high-precision numerical data in others.
The density profile of a cluster with finite mass is found to exhibit exponential decay in $\mathcal{D}=1$ and power-law decay in $\mathcal{D}=2$ with temperature-dependent exponents in both cases.
In $\mathcal{D}=2$ the gas evaporates in a continuous transition at a nonzero critical temperature.
We describe clusters of infinite mass in $\mathcal{D}=3$ with a density profile consisting of three layers (core, shell, halo) and an algebraic large-distance asymptotic decay.
In $\mathcal{D}=3$ a cluster of finite mass can be stabilized at $T>0$ via confinement to a sphere of finite radius.
In some parameter regime, the gas thus enclosed undergoes a discontinuous transition between distinct density profiles.
For the free energy needed to identify the equilibrium state we introduce a construction of gravitational self-energy that works in all $\mathcal{D}$ for the lattice gas.
The decay rate of the density profile of an open cluster is shown to transform via a stretched exponential for $1<\mathcal{D}<2$ whereas it crosses over from one power-law at intermediate distances to a different power-law at larger distances for $2<\mathcal{D}<3$.
\end{abstract}

\maketitle

%
\section{Introduction}\label{sec:intro}
%
This is a statistical mechanical study of a classical gas of massive particles involving short-range repulsive and long-range attractive pair interactions.
The former is a hardcore exclusion interaction and the latter a Newtonian gravitational force analyzed in situations of spherical, cylindrical, and planar symmetry.
The latter two situations are customarily described as modified long-range interactions operating in lower-dimensional spaces.

The interplay between interactions and thermal fluctuations is well known to produce ordering tendencies that strongly depend on dimensionality $\mathcal{D}$.
In cases of interactions that are exclusively of short range, all evidence points to a weakening of fluctuations and a strengthening of ordering tendencies with increasing $\mathcal{D}$. 
Long-range attractive forces reverse the relationship between ordering tendency and dimensionality in at least one sense: the stability of self-gravitating clusters against evaporation decreases as $\mathcal{D}$ increases.

The lattice gas with short-range attractive forces confined to a box is known to undergo a phase transition at temperatures $T>0$ only in $\mathcal{D}\geq2$.
Mean-field predictions for the critical singularities are accurate only at $\mathcal{D}\geq4$ \cite{Fish74,Stan87}.
The self-gravitating lattice gas also features marginal dimensionalities.
In $\mathcal{D}<2$ the gas is stable against evaporation at all finite $T$ and no transitions of any kind occur.
Stable clusters of finite mass at finite $T$ only exist in $\mathcal{D}\leq2$.
Stable clusters in $\mathcal{D}\geq3$ do exist at $T>0$ but have infinite mass.
Thermal fluctuations are reined in by the long-range interactions to render mean-field predictions accurate in all $\mathcal{D}$ with few caveats.

A different but no less vital part of the lattice gas is played by the hardcore exclusion interaction.
It prevents the gas from suffering a gravitational collapse at low $T$, which is well known to happen to a classical gas of point particles \cite{DdV07, dVS06}.
Different schemes \cite{AH72, SKS95, Chav02, IK03, DdV07, CA15, MARF17, FL00} of short-distance regularization have been used before with considerable success and consistency as substitutes for the Pauli principle operating in fermionic matter \cite{Chav02, Chav04}. 

The lattice gas has rarely been invoked for collapse-proof self-gravitating gases.
Notable exceptions are papers by Chavanis \cite{Chav14} and by Pirjol and Schat \cite{PS15}.
The advantages offered by the lattice-gas equation of state include that its structure is simple, fully transparent, microscopically grounded, and independent of $\mathcal{D}$.
Its built-in hardcore repulsion serves the dual purpose of removing short-distance divergences and of providing stability against (artificial) gravitational collapse.
The density profiles of all macrostates that are mechanically and thermally stable can be derived from a single nonlinear second-order ordinary differential equation (ODE) with physically motivated boundary conditions and the two parameters $T$ and $\mathcal{D}$.

The study of self-gravitating gases has a long tradition in statistical physics and astrophysics with an impressive record of findings for stable and metastable states and for processes close to and far from equilibrium \cite{CDS09, LW68, Chand42}.
The topics closest to our work have been admirably reviewed by Chavanis \cite{Chav06} and Padmanabhan \cite{Padm90}.

The inequivalence of statistical ensembles and the validity range of mean-field theory are two aspects that matter for our study but will not be points of emphasis.
They have already been treated rather comprehensively \cite{BMR01, Elli99, BB05, BBDR05, BB06, LP13, CADR14, HT71a, HT71b, dVS02}.
Our work adds to the numerous studies of self-gravitating classical gases new results for the shape and the decay laws of density profiles in open and closed, finite and infinite clusters, at high and low $T$, in $\mathcal{D}$-dimensional space.

Existing results for density profiles pertaining to a gas of classical point particles are readily reproduced in the low-density limit of our analysis.
The lattice gas model at higher densities exhibits signature effects of the hardcore repulsion in the density and pressure profiles.

In Sec.~\ref{sec:equi-cond} we establish the dual conditions of mechanical and thermal equilibrium that constitute the foundation for the statistical mechanical analysis. 
We derive differential equations for the radial profiles of density, pressure, and gravitational potential, including boundary conditions for closed and open systems.
We also construct an expression for the gravitational self-energy that can be used consistently in all $\mathcal{D}$, specifically as part of the free energy needed to identify the equilibrium state among multiple solutions.
In Sec.~\ref{sec:equi-state} we present density and pressure profiles for a closed system of finite mass in $\mathcal{D}=1,2,3$, stabilized into a cluster by gravity alone or assisted by an outer wall. 
Density profiles of an open system with finite or infinite mass are analyzed in Sec.~\ref{sec:open syst}. 

%
\section{Equilibrium Conditions}\label{sec:equi-cond}
%
The foundations of our model and the tools for its analysis are in line with a host of previous work.
Our claim to originality is the lattice-gas context with focus on density profiles aided by an alternative free-energy expression.

\subsection{Thermal equilibrium}\label{sec:therm-eq}
\begin{figure}[htb]
  \begin{center}
 \includegraphics[width=70mm]{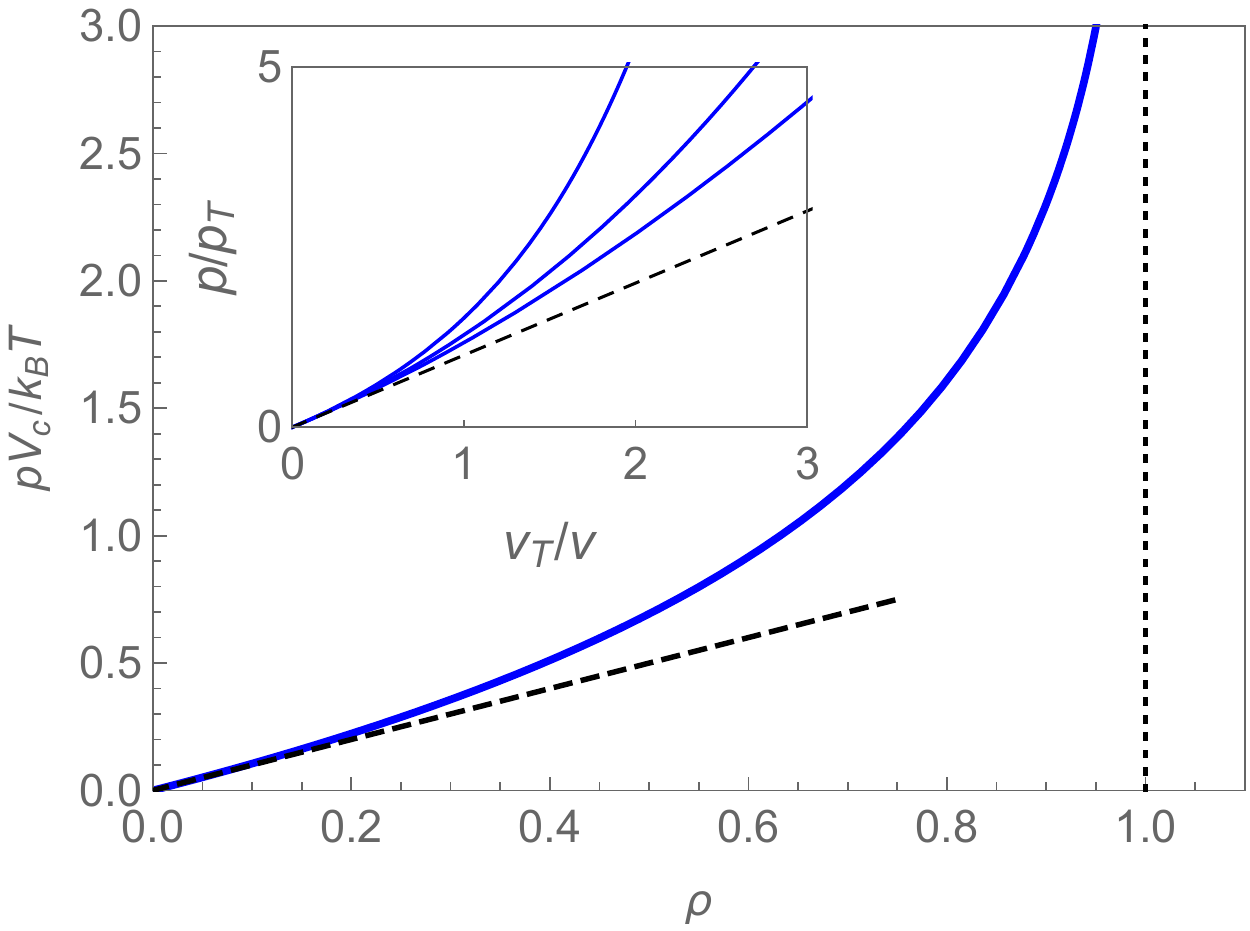}
\end{center}
\caption{Main plot: EOS for the ILG (solid line) and the ICG (dashed line) valid in any $\mathcal{D}$. Inset: Isotherms for the ideal Fermi-Dirac gas in $\mathcal{D}=1,2,3$ (solid lines from top down) with reference values $v_T\doteq \lambda_T^\mathcal{D}$ and $p_T\doteq k_\mathrm{B}T/v_T$ (see Appendix \ref{sec:appe}) and ideal Maxwell-Boltzmann gas (dashed line). The 
curves cross over from a linear, ICG-like behavior at low densities to  a power-law behavior, $\sim (v_T/v)^{1+2/\mathcal{D}}$, at high densities.}
  \label{fig:selgra2t}
\end{figure}

The ideal lattice gas (ILG) in a closed, homogeneous environment consists of $N_\mathrm{c}$ cells of volume $V_\mathrm{c}$ with $N$ particles distributed among them.
The prohibition of multiple cell occupancy represents a hardcore repulsive interaction between particles.
The equation of state (EOS), which expresses the equilibrium relation between the (spatially uniform) intensive state variables pressure $p$, temperature $T$, and density $\rho$, is well known for the ILG and approaches that of the ideal classical gas (ICG) upon dilution \cite{Cha87, Yeo92, sivp, BMM13, inharo}:
\begin{equation}\label{eq:117} 
\frac{pV_\mathrm{c}}{k_\mathrm{B}T}=-\ln(1-\rho) ~\stackrel{\rho\ll1}{\leadsto}~ \rho,\quad \rho=\frac{N}{N_\mathrm{c}}.
\end{equation}
A graphical representation of the EOS for the ILG and its ICG asymptotics is shown in Fig.~\ref{fig:selgra2t} (main plot).
 
The hardcore repulsive interaction of the ILG provides mechanical stability against collapse at high $p$ or low $T$ and approximates (in overly sturdy manner) an effect of the Pauli exclusion principle operating in fermionic matter \cite{Chav06,Chav04}.
For comparison we show in Fig.~\ref{fig:selgra2t} (inset) isotherms of the ideal Fermi-Dirac (FD) gas in dimensions $\mathcal{D}=1,2,3$.

\subsection{Mechanical equilibrium}\label{sec:mech-eq}
In the presence of an external potential $\mathcal{U}(\mathbf{r})$, the thermal equilibrium state is described, at uniform $T$, by profiles $p(\mathbf{r})$ and $\rho(\mathbf{r})$.
The EOS (\ref{eq:117}) still holds locally under mild assumptions.
The local balancing of forces is expressed by an equation of motion (EOM) that relates $\mathcal{U}(\mathbf{r})$ with $p(\mathbf{r})$ and $\rho(\mathbf{r})$.

In the self-gravitating ILG, the potential $\mathcal{U}(\mathbf{r})$ is derived from the interaction potential (energy) between particles of mass $m_\mathrm{c}$ occupying cells a distance $r_{ij}$ apart:
\begin{equation}\label{eq:55} 
\frac{\mathcal{V}_{ij}}{Gm_\mathrm{c}^2}
~~=\left\{
\begin{array}{ll}
~r_{ij} &:~ \mathcal{D}=1,  \rule[-2mm]{0mm}{6mm}\\
~\ln r_{ij} &:~ \mathcal{D}=2,  \rule[-2mm]{0mm}{6mm}\\
-r_{ij}^{-1} &:~ \mathcal{D}=3,  \rule[-2mm]{0mm}{6mm}
\end{array}\right.
\end{equation}
where $G$ is a ($\mathcal{D}$-dependent) constant of gravitation.
The gravitational interaction force,
\begin{equation}\label{eq:118}
F_{ij}=-\frac{Gm_\mathrm{c}^2}{r_{ij}^{\mathcal{D}-1}},
\end{equation}
obeys the familiar inverse-square law in $\mathcal{D}=3$ and has been generalized  to satisfy Gauss' law also in $\mathcal{D}=1,2$.

In a radially symmetric self-gravitating cluster with center of mass at $\mathbf{r}=0$, Gauss' law for the gravitational field $g(r)$ or potential $\mathcal{U}(r)$ reduces to
\begin{equation}\label{eq:56} 
g(r)\doteq-\frac{d\mathcal{U}}{dr}=-\frac{Gm_\mathrm{in}}{r^{\mathcal{D}-1}},
\end{equation}
where $m_\mathrm{in}$ is the mass of all occupied cells inside radius $r$ and related to the density profile $\rho(r)$ as follows:
\begin{equation}\label{eq:119} 
m_\mathrm{in}=\frac{m_\mathrm{c}}{V_\mathrm{c}}\int_0^r dr'\big(A_\mathcal{D}r'^{\mathcal{D}-1}\big)\rho(r'),
\end{equation}
where
\begin{equation}\label{eq:59} 
\mathcal{A}_\mathcal{D}=\frac{2\pi^{\mathcal{D}/2}}{\Gamma(\mathcal{D}/2)}
=\left\{\begin{array}{ll}
2 &: \mathcal{D}=1, \rule[-2mm]{0mm}{5mm}\\ 2\pi &: \mathcal{D}=2, \rule[-2mm]{0mm}{5mm}\\ 4\pi &: \mathcal{D}=3, \rule[-2mm]{0mm}{5mm}
\end{array} \right.
\end{equation}
is the surface area of the $\mathcal{D}$-dimensional unit sphere.

The two conditions of thermal equilibrium (EOS) and hydrostatic equilibrium (EOM),
\begin{equation}\label{eq:120} 
\frac{p(r)V_\mathrm{c}}{k_\mathrm{B}T}=-\ln\big(1-\rho(r)\big),
\end{equation}
\begin{equation}\label{eq:121} 
\frac{d}{dr}p(r)=\frac{m_\mathrm{c}}{V_\mathrm{c}}\rho(r)g(r),
\end{equation}
respectively, the latter in conjunction with (\ref{eq:56}) and (\ref{eq:119}), constitute a closed set of relations between the functions $\rho(r)$, $p(r)$, and $\mathcal{U}(r)$ at given (uniform) $T$.

\subsection{Differential equations}\label{sec:dif-equ}
For the purpose of our analysis it is convenient to use the dimensionless scaled variables,
\begin{equation}\label{eq:2} 
\hat{r}\doteq\frac{r}{r_\mathrm{s}},\quad \hat{p}\doteq\frac{p}{p_\mathrm{s}},\quad 
\hat{T}\doteq\frac{k_\mathrm{B}T}{p_\mathrm{s}V_\mathrm{c}},\quad 
\hat{\mathcal{U}}\doteq\frac{\mathcal{U}m_\mathrm{c}}{p_\mathrm{s}V_\mathrm{c}},
\end{equation}
for radius, pressure, temperature, and potential, respectively, with reference values
\begin{equation}\label{eq:3} 
 r_\mathrm{s}^\mathcal{D}
=\frac{NV_\mathrm{c}\mathcal{D}}{\mathcal{A_D}},\quad 
p_\mathrm{s}
=\frac{\mathcal{A}_\mathcal{D}G}{2\mathcal{D}}\frac{m_\mathrm{c}^2}{V_\mathrm{c}^2}r_\mathrm{s}^2.
\end{equation}

In the analysis at $T>0$ we express the EOS (\ref{eq:120}) and the EOM (\ref{eq:121}) with (\ref{eq:56}) and (\ref{eq:119}) using these scaled variables,
\begin{equation}\label{eq:60} 
\hat{p}(\hat{r})=-\hat{T}\ln\big(1-\rho(\hat{r})\big),
\end{equation}
\begin{equation}\label{eq:62} 
\frac{d\hat{p}}{d\hat{r}}=-\rho(\hat{r})\frac{d\hat{\mathcal{U}}}{d\hat{r}}
=-2\mathcal{D}\rho(\hat{r})\int_0^{\hat{r}}d\hat{r}'\rho(\hat{r}')
\left(\frac{\hat{r}'}{\hat{r}}\right)^{\mathcal{D}-1},
\end{equation}
and infer the relation,
\begin{equation}\label{eq:88} 
\hat{\mathcal{U}}(\hat{r})=\hat{T}\ln\left(\frac{1-\rho(\hat{r})}{1-\rho(0)}\,\frac{\rho(0)}{\rho(\hat{r})}\right),
\end{equation}
between potential and density with the (convenient) reference value $\hat{\mathcal{U}}(0)=0$ imposed.
Elimination of $\hat{p}(\hat{r})$ yields
\begin{equation}\label{eq:125}
\frac{\hat{T}\rho'(\hat{r})}{\rho(\hat{r})[1-\rho(\hat{r})]}
=-2\mathcal{D}\int_0^{\hat{r}} d\hat{r}'\left(\frac{\hat{r}'}{\hat{r}}\right)^{\mathcal{D}-1}\rho(\hat{r}'),
\end{equation}
from which we conclude that the density must be a monotonically decreasing function of $\hat{r}$ with zero initial slope, $\rho'(0)=0$.
Equation (\ref{eq:60}) then implies that $\hat{p}'(0)=0$.

It is useful to convert (\ref{eq:125}) into the second-order nonlinear ODE for the density profile,
\begin{equation}\label{eq:63} 
\frac{\rho''}{\rho} +\frac{\mathcal{D}-1}{\hat{r}}\frac{\rho'}{\rho} -\frac{1-2\rho}{1-\rho}\left(\frac{\rho'}{\rho}\right)^2 
+\frac{2\mathcal{D}}{\hat{T}}\rho(1-\rho)=0.
\end{equation}
Equivalent ODEs for pressure and potential,
\begin{equation}\label{eq:64} 
\hat{p}''+\frac{\mathcal{D}-1}{\hat{r}}\hat{p}'-\frac{1-\rho}{\hat{T}\rho}\hat{p}'^2
+2\mathcal{D}\rho^2=0,
\end{equation}
\begin{equation}\label{eq:76} 
\hat{\mathcal{U}}''+\frac{\mathcal{D}-1}{\hat{r}}\,\hat{\mathcal{U}}' -2\mathcal{D}\rho=0,
\end{equation}
imply the use of (\ref{eq:60}) and (\ref{eq:88}) if $\hat{T}>0$ \cite{note3}.
These last two ODEs are most often used in the limit $\hat{T}\to0$, where the functional relations (\ref{eq:60}) and (\ref{eq:88}) break down.

The ODEs (\ref{eq:63})-(\ref{eq:76}) also hold in open systems.
Here (\ref{eq:88}) is best rendered in the form
\begin{equation}\label{eq:116}
 \rho(\hat{r})=\frac{1}{1+e^{(\hat{\mathcal{U}}-\hat{\mu})/\hat{T}}},
\end{equation}
where the (scaled) chemical potential,
\begin{equation}\label{eq:115} 
 \hat{\mu}\doteq\frac{\mu}{p_\mathrm{s}V_\mathrm{c}}
 =-\hat{T}\ln\left(\frac{1-\rho(0)}{\rho(0)}\right),
\end{equation}
controls the average number of particles.

\subsection{Boundary conditions}\label{sec:boun-cond}
The physically relevant boundary conditions of (\ref{eq:63}) or (\ref{eq:64}) for a closed system (fixed $N$) confined to a region of maximum radius $\hat{R}\doteq R/r_\mathrm{s}>1$ involve one local relation,
\begin{equation}\label{eq:65} 
\rho'(0)=0,\quad \hat{p}'(0)=0,
\end{equation}
and one nonlocal relation for $\rho(0)=\rho_0$, $\hat{p}(0)=\hat{p}_0$, namely
\begin{equation}\label{eq:67} 
\mathcal{D}\int_0^{\hat{R}} d\hat{r}\,\hat{r}^{\mathcal{D}-1}\rho(\hat{r})=1,
\end{equation}
\begin{subequations}\label{eq:66} 
\begin{align}\label{eq:66a} 
\hat{p}(0)=1+\hat{p}(\hat{R}) &:~ \mathcal{D}=1 \\
2\mathcal{D}(\mathcal{D}-1)\int_0^{\hat{R}} d\hat{r}\,\hat{r}^{2\mathcal{D}-3}\hat{p}(\hat{r})
\hspace*{10mm}& \nonumber \\ \label{eq:66b} 
=1+\mathcal{D}\hat{R}^{2(\mathcal{D}-1)}\hat{p}(\hat{R}) &:~ \mathcal{D}>1,
\end{align}
\end{subequations}
respectively.
On some occasions, the integral conditions have multiple solutions for a given $\rho_0$ or $\hat{p}_0$.
In one such case (Sec.~\ref{sec:Deq3}), three solutions are identified as representing a stable, a metastable, and an unstable density profile.

The local conditions (\ref{eq:65}) follow from Eqs.~(\ref{eq:60}) and (\ref{eq:62}) as discussed earlier.
The nonlocal condition (\ref{eq:67}) reflects particle conservation and (\ref{eq:66}) is derived from integration of (\ref{eq:62}).
In the absence of wall confinement we set $\hat{p}(\hat{R})=0$ for $\hat{R}\to\infty$.
In $\mathcal{D}=1$, where the interaction force (\ref{eq:118}) is independent of distance, the pressure at the center of an unconfined cluster is invariant: $\hat{p}(0)=1$.
Both boundary conditions of (\ref{eq:76}) are local,
\begin{equation}\label{eq:114} 
 \hat{\mathcal{U}}(0)=\hat{\mathcal{U}}'(0)=0,
\end{equation}
and follow from Eqs.~(\ref{eq:62}) and (\ref{eq:88}).

Using the center of a symmetric cluster as the reference point for the potential differs from common practice in Newtonian mechanics $(\mathcal{D}=3)$ but is more convenient for comparisons with results in $\mathcal{D}=1,2$.
We then have $\mathcal{U}(\hat{r})\geq0$ at any radius.

In open systems, conditions (\ref{eq:65}) still hold, whereas (\ref{eq:67}) needs to be replaced by $\rho(0)=1/(1+e^{-\hat{\mu}/\hat{T}})$, and (\ref{eq:66}) by the value for $\hat{p}(0)$ inferred from $\rho(0)$ via (\ref{eq:60}).

\subsection{ICG limit}\label{sec:Asym}
If we use the EOS of the ICG, $\hat{p}(\hat{r})=\hat{T}\rho(\hat{r})$, instead of the EOS (\ref{eq:60}) of the ILG in the transformations of Sec.~\ref{sec:dif-equ} we end up with the ODE,
\begin{equation}\label{eq:68} 
\frac{\rho''}{\rho}+\frac{\mathcal{D}-1}{\hat{r}}\frac{\rho'}{\rho}-\left(\frac{\rho'}{\rho}\right)^2
+\frac{2\mathcal{D}}{\hat{T}}\rho=0,
\end{equation}
which is a low-density approximation of (\ref{eq:63}).
The effects of hardcore repulsion are no longer present.
This ODE for $\mathcal{D}=3$ is well known in astrophysics as a Lane-Emden type equation \cite{Emde07}. 
The solutions of (\ref{eq:68}) are relevant for the ILG in regimes where $\rho(\hat{r})\ll1$ holds. 
This can be the case locally at large $\hat{r}$ or globally at high $\hat{T}$.

It is worthwhile to discuss the ICG density profiles in some detail.
They exhibit attributes of universality which their ILG counterparts do not.
These features of universality are best brought into focus if we introduce further sets of scaled variables.

(i) For a closed ICG system (of finite mass) confined to a space of maximum radius $\hat{R}$ we set
\begin{equation}\label{eq:131} 
\bar{r}\doteq\frac{\hat{r}}{\hat{R}},\quad \bar{\rho}\doteq\hat{R}^\mathcal{D}\rho,\quad 
\bar{T}\doteq\hat{R}^{\mathcal{D}-2}\hat{T},
\end{equation}
which leaves the structure of (\ref{eq:68}) invariant,
\begin{equation}\label{eq:132} 
\frac{\bar{\rho}''}{\bar{\rho}}+\frac{\mathcal{D}-1}{\bar{r}}\frac{\bar{\rho}'}{\bar{\rho}}
-\left(\frac{\bar{\rho}'}{\bar{\rho}}\right)^2
+\frac{2\mathcal{D}}{\bar{T}}\bar{\rho}=0,
\end{equation}
and removes the $\hat{R}$-dependence from the condition (\ref{eq:67}):
\begin{equation}\label{eq:133} 
\mathcal{D}\int_0^{1} d\bar{r}\,\bar{r}^{\mathcal{D}-1}\bar{\rho}(\bar{r})=1.
\end{equation}

(ii) For an open cluster (of finite or infinite mass) stabilized by gravity alone we set [with $\rho_0=\rho(0)$]
\begin{equation}\label{eq:134} 
\tilde{r}\doteq\sqrt{\frac{2\mathcal{D}\rho_0}{\hat{T}}}\;\hat{r},\quad \tilde{\rho}\doteq\frac{\rho}{\rho_0}.
\end{equation}
This choice produces the ODE,  
\begin{equation}\label{eq:135} 
\frac{\tilde{\rho}''}{\tilde{\rho}}+\frac{\mathcal{D}-1}{\tilde{r}}\frac{\tilde{\rho}'}{\tilde{\rho}}
-\left(\frac{\tilde{\rho}'}{\tilde{\rho}}\right)^2
+ \tilde{\rho}=0,
\end{equation}
with (local) boundary conditions,
\begin{equation}\label{eq:136} 
\tilde{\rho}(0)=1,\quad \tilde{\rho}'(0)=0.
\end{equation}

Both rescaling operations (i) and (ii) provide useful low-density benchmarks for the ILG.

\subsection{Free energy}\label{sec:free-ener}
In situations where Eq.~(\ref{eq:63}) admits multiple solutions for physically relevant boundary conditions, the equilibrium state will be represented by the solution with the lowest free energy. 
For a closed system with a finite number $N$ of particles stabilized by gravity alone or assisted by a rigid wall at radius $\hat{R}$, 
the relevant thermodynamic potential is the (dimensionless) Helmholtz free energy,
\begin{equation}\label{eq:111}
\hat{\mathcal{F}}(\hat{T})=\hat{U}_\mathrm{S}-\hat{T}\hat{\mathcal{S}}.
\end{equation}
$\hat{U}_\mathrm{S}\doteq U_\mathrm{S}/Np_\mathrm{s}V_\mathrm{c}$ is the  gravitational self-energy relative to a reference state of choice.
$\hat{\mathcal{S}}$ is the ILG entropy density, e.g. from \cite{sivp}, integrated over the space available to the particles:
\begin{subequations}\label{eq:126}
\begin{equation}\label{eq:126a} 
 \hat{\mathcal{S}}\doteq \frac{\mathcal{S}}{Nk_\mathrm{B}}
= \mathcal{D}\int_0^{\hat{R}} d\hat{r}\,\hat{r}^{\mathcal{D}-1}\bar{S}[\rho],
\end{equation}
\begin{align}\label{eq:126b} 
\bar{S}[\rho] =-\rho\ln\rho-(1-\rho)\ln(1-\rho),
\end{align}
\end{subequations}
with $\hat{R}\to\infty$ in the absence of wall confinement.

The construction of $\hat{U}_\mathrm{S}$ in $\mathcal{D}$ dimensions requires circumspection.
The commonly used expression of gravitational self-energy $U_\mathrm{S}^{(\mathrm{F})}$ for a symmetric cluster in $\mathcal{D}=3$ is the quantity $\frac{1}{2}\rho_\mathrm{m}(r)\mathcal{U}_\mathrm{F}(r)$ integrated over the (finite or infinite) space occupied by the cluster.
Here $\rho_\mathrm{m}(r)$ is the mass density and $\mathcal{U}_\mathrm{F}(r)=-Gm_\mathrm{in}r^{-1}$ the gravitational potential generated by the (symmetric) cluster.
With the convention $\mathcal{U}_\mathrm{F}(\infty)=0$, the (negative) self-energy $U_\mathrm{S}^{(\mathrm{F})}$ thus obtained can be interpreted as the change in potential energy during the assembly of a cluster of particles that originate from places out at infinity, where their interaction potential (\ref{eq:55}) vanishes.
The trouble is that in $\mathcal{D}\leq2$ there are no such locations.

The only reference point for the gravitational potential that is practical in all $\mathcal{D}$ is at the center of the cluster: $\mathcal{U}(0)=0$.
A practical reference value for the self-energy then also depends on a convenient reference configuration of particles.
For a finite ILG cluster (closed system) the obvious reference configuration is the ground state, a symmetric cluster of unit density for $0\leq r\leq r_\mathrm{s}$ as described below in Sec.~\ref{sec:Teq0}.
The gravitational self-energy $U_\mathrm{S}$ of any other macrostate relative to the ground state is then positive.

In Appendix~\ref{sec:appc} we derive an integral expression for $U_\mathrm{S}$ that works in any dimension $\mathcal{D}\geq1$.
We also prove the equality, $\Delta U_\mathrm{S}=\Delta U_\mathrm{S}^{(\mathrm{F})}$, in $\mathcal{D}=3$ between macrostates with arbitrary density profiles.
The scaled self-energy expression reads
\begin{equation}\label{eq:113} 
\hat{U}_\mathrm{S}=\left\{
\begin{array}{l}
{\displaystyle \frac{2\mathcal{D}}{\mathcal{D}-2}\int_0^\infty 
d\hat{r}_2\,\rho(\hat{r}_2)
\Big[\hat{r}_1^2\hat{r}_2^{\mathcal{D}-1}-\hat{r}_1^\mathcal{D}\hat{r}_2\Big]}, 
\rule[-2mm]{0mm}{10mm} \\ \rule[-2mm]{0mm}{10mm} 
{\displaystyle 4\int_0^\infty
d\hat{r}_2\,\rho(\hat{r}_2)\hat{r}_1^2\hat{r}_2\ln\frac{\hat{r}_2}{\hat{r}_1}},
\end{array} \right.
\end{equation}
for $\mathcal{D}\neq2$ and $\mathcal{D}=2$, respectively,
where $\hat{r}_1$ depends on the integration variable $\hat{r}_2$ via
\begin{equation}\label{eq:112} 
\hat{r}_1^\mathcal{D}=\mathcal{D}\int_0^{\hat{r}_2}d\hat{r}\,\hat{r}^{\mathcal{D}-1}\rho(\hat{r}).
\end{equation}

%
\section{Closed systems}\label{sec:equi-state}
%
Here we present density profiles (and some pressure profiles) for self-gravitating ILG clusters that are closed in the thermodynamic sense.
The accessible space is infinite in some cases and finite in others.
The total mass is finite and fixed in all cases.

\subsection{$\hat{T}=0$}\label{sec:Teq0}
At zero temperature the ILG forms a solid cluster of radius $r_\mathrm{s}$ containing $N$ particles.
The density has a step discontinuity,
\begin{equation}\label{eq:71} 
\rho(\hat{r})=\theta(1-\hat{r}).
\end{equation}
The pressure profile inferred from (\ref{eq:62}) is quadratic,
\begin{equation}\label{eq:61} 
\hat{p}(\hat{r})=\big(1-\hat{r}^2\big)\theta(1-\hat{r}),
\end{equation}
with reference pressure $p_\mathrm{s}$ realized at $r=0$.
The ODE (\ref{eq:63}) reduces to $\rho(1-\rho)=0$, which is consistent with (\ref{eq:71}), and the ODE (\ref{eq:64}) to $\hat{p}''+{(\mathcal{D}-1)}\hat{p}'/\hat{r}+2\mathcal{D}=0$ for $\hat{r}\leq1$, which is consistent with (\ref{eq:61}).

\begin{figure}[t]
  \begin{center}
 \includegraphics[width=70mm]{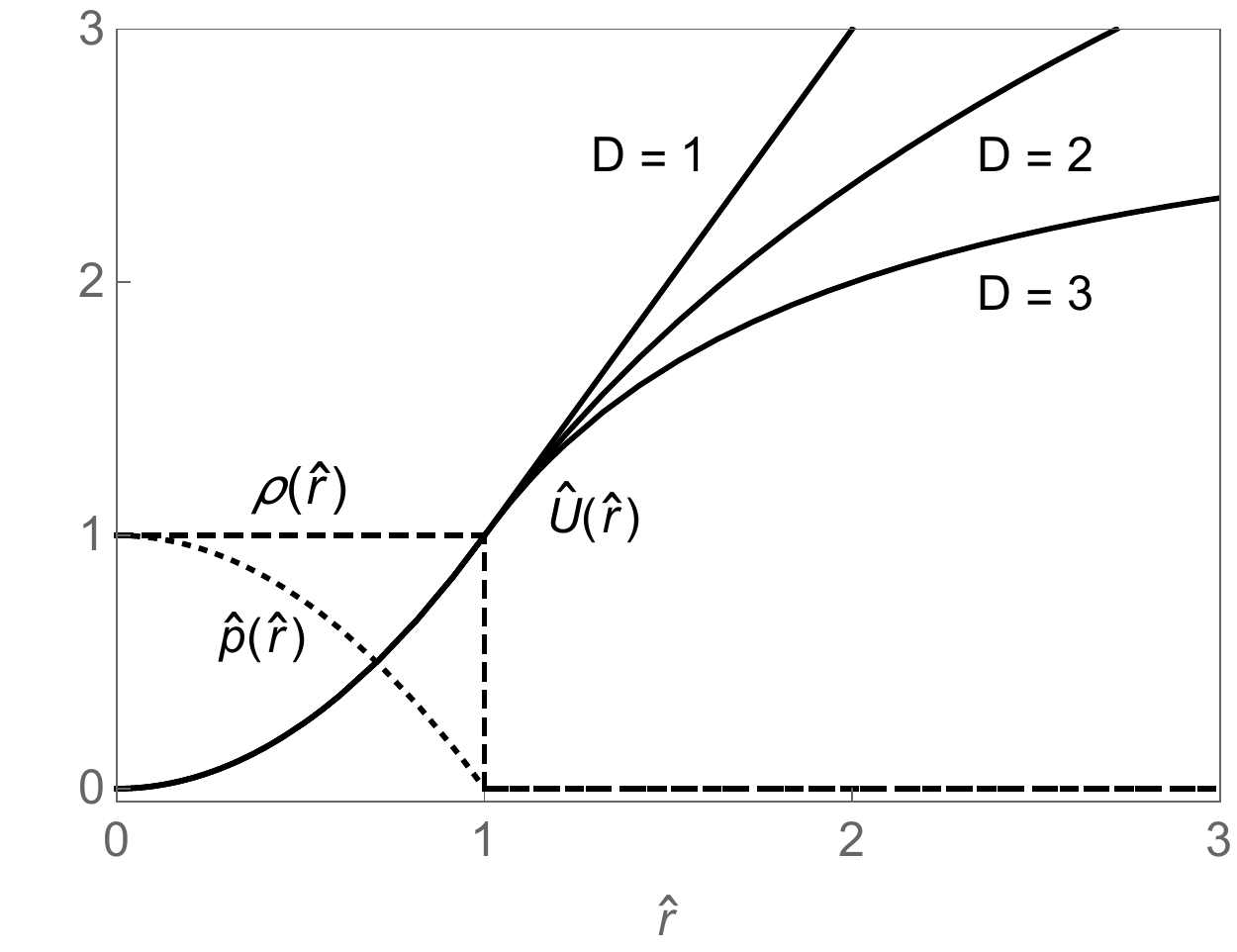}
\end{center}
\caption{Scaled density, pressure, and potential versus scaled radius of a self-gravitating ILG in $\mathcal{D}=1,2,3$ at $\hat{T}=0$.}
  \label{fig:fig4}
\end{figure}

For the potential we solve (\ref{eq:76}) with (\ref{eq:114}) and use $\rho(\hat{r})$ from (\ref{eq:71}).
The resulting expression in scaled units (\ref{eq:2}) is
\begin{subequations}\label{eq:78}
\begin{equation}\label{eq:78a} 
\hat{\mathcal{U}}(\hat{r})=\hat{r}^2\quad:~ 0\leq\hat{r}\leq1,
\end{equation}
\begin{equation}\label{eq:78b} 
\hat{\mathcal{U}}(\hat{r})=\left\{\begin{array}{ll}
2\hat{r}-1 &:~ \mathcal{D}=1, \\
2\ln\hat{r}+1 &:~ \mathcal{D}=2, \\
3-2/\hat{r} &:~ \mathcal{D}=3, 
\end{array} \right.\quad \hat{r}\geq 1.
\end{equation}
\end{subequations}

At large distances, $\hat{\mathcal{U}}(\hat{r})$ rises to infinity linearly in $\mathcal{D}=1$ and logarithmically in $\mathcal{D}=2$, but levels off to a finite value in $\mathcal{D}=3$.
In Fig.~\ref{fig:fig4} we show the $\hat{T}=0$ profiles (\ref{eq:71}), (\ref{eq:61}), and (\ref{eq:78}) in a comparative plot.
It is well-known that any finite cluster is unstable against evaporation in $\mathcal{D}>2$ due to the non-confining nature of the gravitational attraction.

\subsection{$\mathcal{D}=1$}\label{sec:Deq1}
\begin{figure}[htb]
  \begin{center}
 \includegraphics[width=43mm]{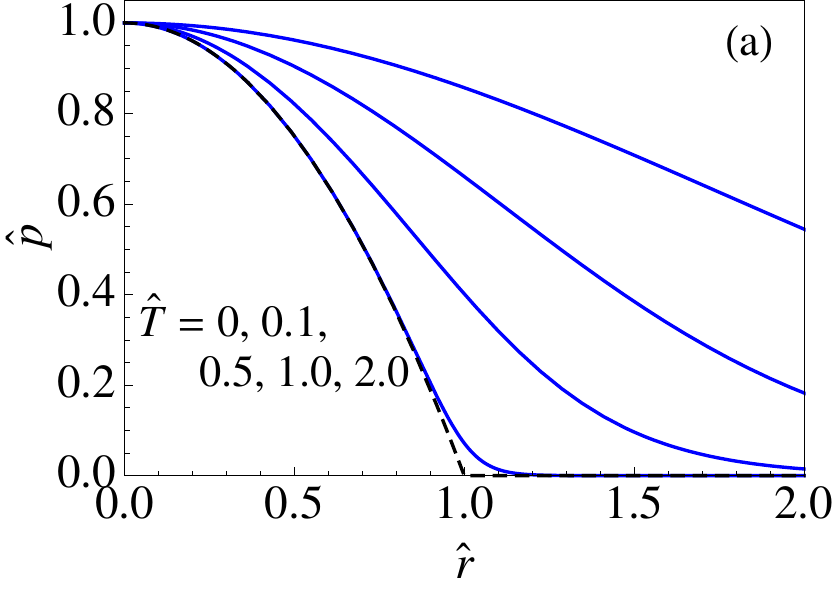}
 \includegraphics[width=43mm]{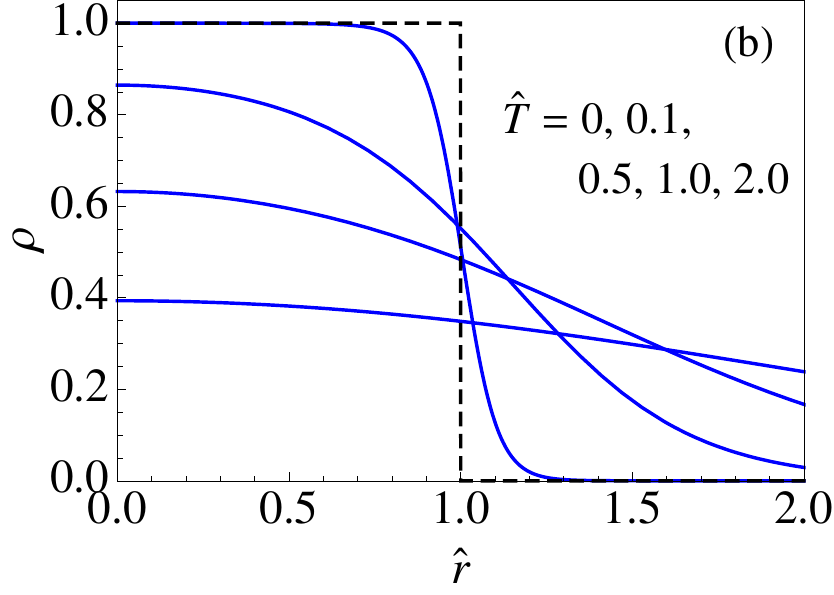}
 
 \includegraphics[width=44mm]{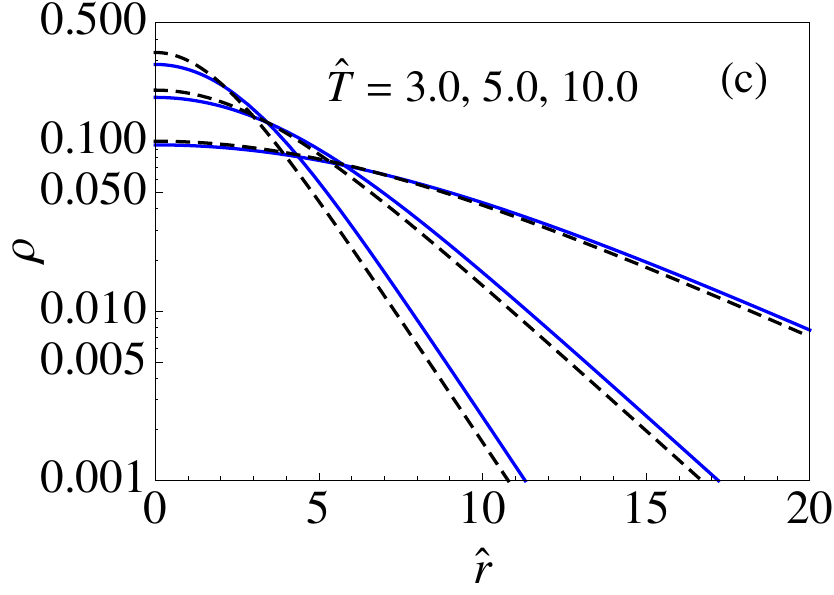}
 \includegraphics[width=42mm]{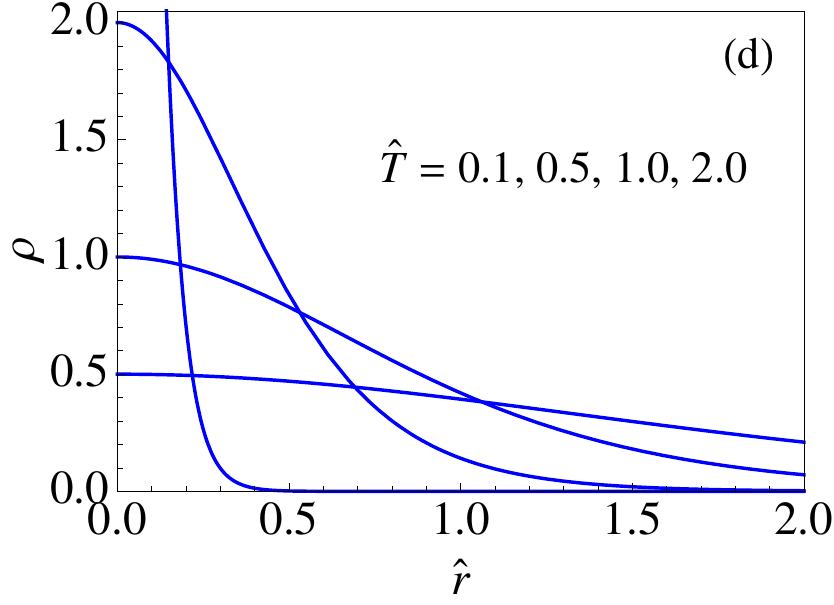}
\end{center}
\caption{Profiles in $\mathcal{D}=1$ of (a) pressure and (b) density for the self-gravitating ILG cluster at $\hat{T}=0$ (dashed curve) and $\hat{T}>0$ (solid curves). (c) ILG density profile at higher $\hat{T}$ (solid curves) in comparison with the asymptotic ICG profiles (\ref{eq:69}) (dashed curves) in a log plot. (d) ICG density profiles (\ref{eq:69}) at low $\hat{T}$.}
  \label{fig:fig6}
\end{figure}

The solution of the ODE (\ref{eq:63}) in $\mathcal{D}=1$ with $\hat{R}=\infty$ produces the curves depicted in Fig.~\ref{fig:fig6}.
Increasing $\hat{T}$ from zero converts the sharp solid surface at $\hat{r}=1$ into an interface of increasing width between 
a high-density core at $\hat{r}<1$ sandwiched between low-density wings at $\hat{r}>1$.
The density profile softens and broadens but the cluster stays intact at any finite $\hat{T}$.
Near the center of the cluster $\rho$ decreases as the gas spreads out [Fig.~\ref{fig:fig6}(b)].
The pressure is invariant at the center of the cluster, $\hat{p}(0)=1$ as explained in Sec.~\ref{sec:boun-cond}, everywhere else it increases as $\hat{T}$ rises.
The pressure profile remains monotonically decreasing but becomes increasingly flat [Fig.~\ref{fig:fig6}(a)].

The exact asymptotic behavior of the ILG density profile is an exponential decay with  $\hat{T}$-dependent exponent,
\begin{equation}\label{eq:85} 
\rho(\hat{r})_\mathrm{as}\sim e^{-2\hat{r}/\hat{T}}\quad:~ \hat{r}\geq1,
\end{equation}
as proven in Appendix \ref{sec:appa}.
It is consistent with the analytic solution,
\begin{equation}\label{eq:69} 
\rho(\hat{r})_\mathrm{ICG}=\frac{1}{\hat{T}}\,\mathrm{sech}^2\!\left(\frac{\hat{r}}{\hat{T}}\right),
\end{equation}
of the ODE (\ref{eq:68}) representing the ICG.

In Fig.~\ref{fig:fig6}(c) we compare the numerical ILG solutions with the analytic ICG solution (\ref{eq:69}).
At all three values of $\hat{T}$ the rate of exponential tailing off agrees.
With increasing $\hat{T}$ the agreement improves overall.
The ICG result (\ref{eq:69}) was found previously and used in a variety of physics contexts \cite{Rybi71, SC02, CBM+13}.

The asymptotic decay (\ref{eq:85}) also emerges from the low-$\hat{T}$ solid-gas approximation invoked in several studies (see Appendix~\ref{sec:appd}). 
Moreover, the density profile (\ref{eq:69}) accurately describes self-gravitating quantum gases (fermions or bosons) at sufficiently low density \cite{Chav04, IR88}.

We note that for the ICG the density profile (\ref{eq:69}) is valid at all $\hat{T}$.
Point particles experience no hardcore repulsion, which permits the density at $\hat{r}=0$ to grow without limit as $\hat{T}\to0$ [Fig.~\ref{fig:fig6}(d)].
However, unlike in higher $\mathcal{D}$, no gravitational collapse at $\hat{T}>0$ takes place.
In $\mathcal{D}=1$ the gravitational force (\ref{eq:118})  does not diverge for $r_{ij}\to0$. 
Confinement by an outer wall at $\hat{R}<\infty$ leaves the ICG density profile (\ref{eq:69}) largely intact.
The solution of (\ref{eq:131})-(\ref{eq:133}) yields
\begin{equation}\label{eq:137} 
\bar{\rho}(\bar{r})_\mathrm{ICG}
=\frac{b\,\mathrm{sech}^2(b\bar{r})}{\mathrm{tanh}\,b},\quad 
b\bar{T}\,\mathrm{tanh}\,b=1.
\end{equation}

\subsection{$\mathcal{D}=2$}\label{sec:Deq2}
\begin{figure}[htb]
  \begin{center}
 \includegraphics[width=43mm]{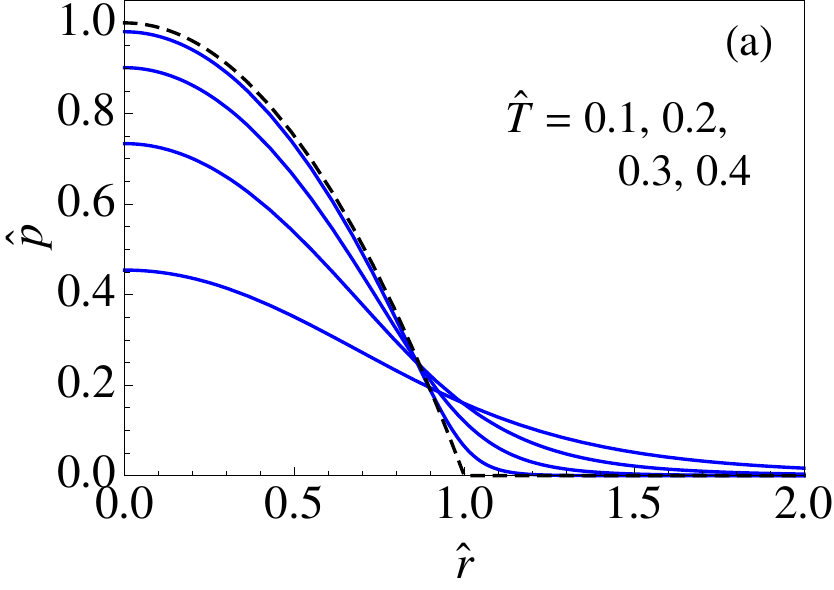}
 \includegraphics[width=43mm]{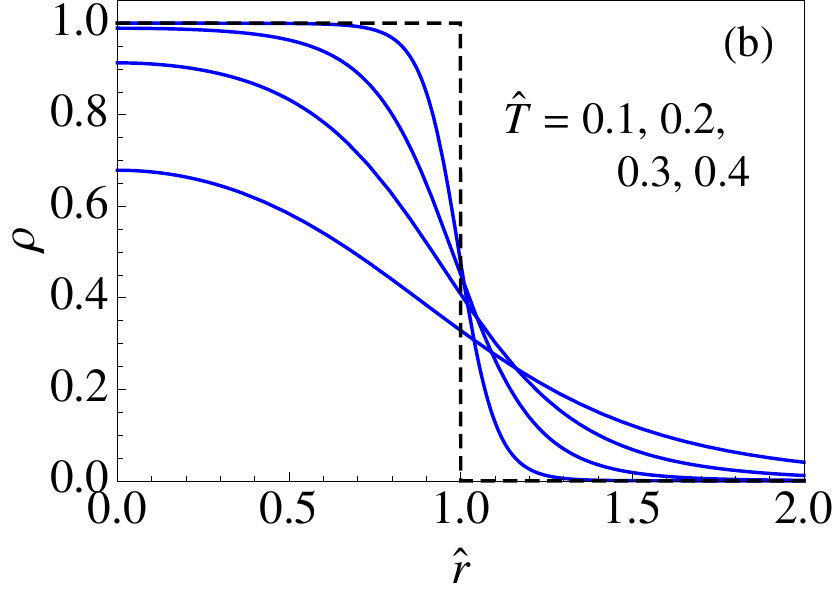}
  
\includegraphics[width=43mm]{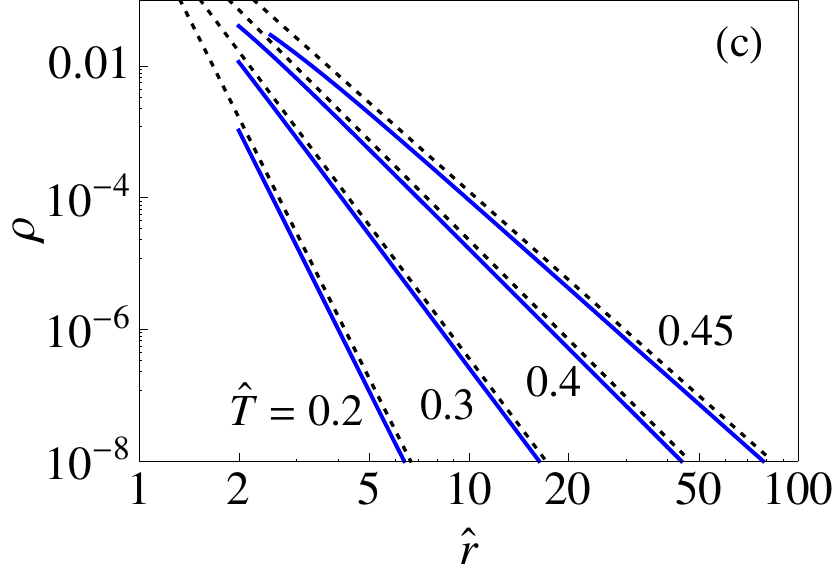}
 \includegraphics[width=43mm]{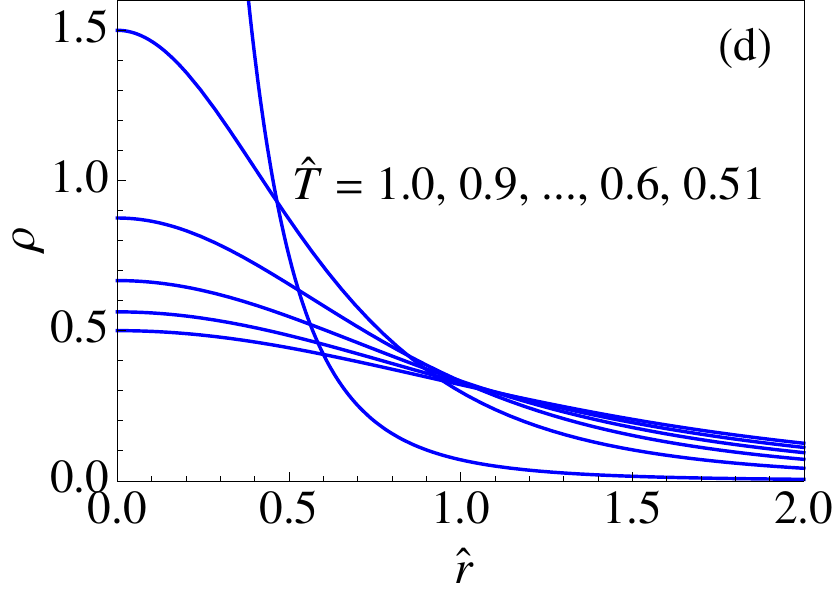}
\end{center}
\caption{Profiles in $\mathcal{D}=2$ of (a) pressure and (b) density for the self-gravitating ILG cluster at $\hat{T}=0$ (dashed curve) and $0<\hat{T}<\hat{T}_\mathrm{c}$ (solid curves). (c) ILG density profiles at $0<\hat{T}<\hat{T}_\mathrm{c}$ (solid curves) in comparison with the power-law asymptotics $\sim\hat{r}^{-2/\hat{T}}$ (dotted lines) in a log-log plot. (d) ICG density profile (\ref{eq:150}) at $\hat{T}>\hat{T}_\mathrm{c}$.}
  \label{fig:fig7}
\end{figure}

The numerical analysis of the ODE (\ref{eq:63}) in $\mathcal{D}=2$ for a finite-mass system with $\hat{R}=\infty$ yields the pressure and density profiles shown in Figs.~\ref{fig:fig7}(a), (b).
Starting from $\hat{T}=0$ (dashed lines) we observe that the pressure at the center of the cluster drops rapidly with rising $\hat{T}$, unlike in $\mathcal{D}=1$.
The density near $\hat{r}=0$ drops more rapidly than it does in $\mathcal{D}=1$.

The power-law decay with $\hat{T}$-dependent exponent of the density is illustrated in Fig.~\ref{fig:fig7}(c).
This numerical evidence is confirmed by the exact leading term,
\begin{equation}\label{eq:82} 
\rho(\hat{r})_\mathrm{as}\sim \hat{r}^{-2/\hat{T}} \quad :~\hat{r}\gg1,
\end{equation}
of the asymptotic behavior as proven in Appendix \ref{sec:appa}.
The solid-gas approximation of Appendix~\ref{sec:appd} predicts the decay law (\ref{eq:82}) to hold throughout the gas albeit with no hint of the impending qualitative changes at higher $\hat{T}$ or smaller $\hat{r}$.

Unlike in $\mathcal{D}=1$, a cluster of finite mass only survives at sufficiently low $\hat{T}$.
The numerical analysis of (\ref{eq:63}) indicates that the density maximum $\rho(0)$ decreases gradually with increasing $\hat{T}$, reaching zero at a finite $\hat{T}_\mathrm{c}$, thus suggesting that the gas evaporates in a continuous transition.
The transition temperature can be pinned down in the ICG limit, which remains accurate for the ILG because evaporation takes place at low density.

We again find an analytic solution of the ODE (\ref{eq:68}) for the ICG but in $\mathcal{D}=2$ this plays out differently. 
Under confinement  $(\hat{R}<\infty)$ and for temperatures exceeding the threshold value,
\begin{equation}\label{eq:144}
\hat{T}_\mathrm{c}=\frac{1}{2},
\end{equation}
the ODE (\ref{eq:68}) produces the exact solution,
\begin{equation}\label{eq:150} 
 \rho(\hat{r})_\mathrm{ICG}=\frac{2\hat{T}}{\hat{R}^2(2\hat{T}-1)}
 \left[1+\frac{1}{2\hat{T}-1}\left(\frac{\hat{r}}{\hat{R}}\right)^2\right]^{-2}.
\end{equation}
For comparison with the $\mathcal{D}=1$ ICG density profile (\ref{eq:69}) plotted in Fig.~\ref{fig:fig6}(d) we show in Fig.~\ref{fig:fig7}(d) the $\mathcal{D}=2$ profile (\ref{eq:150}) for various $\hat{T}>\hat{T}_\mathrm{c}$.
This profile is unstable against gravitational collapse as $\hat{T}$ is lowered past the value $\hat{T}_\mathrm{c}=\frac{1}{2}$.

With scaled variables (\ref{eq:131}) only the parameter $\hat{T}=\bar{T}$ remains:
\begin{equation}\label{eq:130} 
\bar{\rho}(\bar{r})_\mathrm{ICG}=
\frac{2\bar{T}(2\bar{T}-1)}{[\bar{r}^2+2\bar{T}-1]^2}\quad :~ 0\leq\bar{r}\leq1.
\end{equation}
This scaled ICG density profile shares with its $\mathcal{D}=1$ counterpart (\ref{eq:137}) the property of gradually turning into a $\delta$-function at $\bar{r}=0$.
In $\mathcal{D}=1$ this happens at $\bar{T}=0$, in $\mathcal{D}=2$ at $\bar{T}=\bar{T}_\mathrm{c}=\frac{1}{2}$.
The pressure against the outer wall at $\bar{r}=1$ then vanishes in both cases.
The pressure at the center of the ICG cluster stays finite as $\bar{T}\to0$ in $\mathcal{D}=1$ whereas it diverges as $\bar{T}\to\bar{T}_\mathrm{c}$ in $\mathcal{D}=2$.

Returning to scaled variables (\ref{eq:2}), we find that at ${\hat{T}>\hat{T}_\mathrm{c}}$, confinement is necessary to prevent the ICG from evaporating.
If we take the limit $\hat{R}\to\infty$ at $\hat{T}>\hat{T}_\mathrm{c}$, the profile (\ref{eq:150}) flattens and approaches zero.
However, if we take the combined limit,
\begin{equation}\label{eq:151} 
  \hat{T}\to\hat{T}_\mathrm{c},\quad \hat{R}\to\infty,\quad
  \frac{\hat{T}^2}{2\hat{R}^2(2\hat{T}-1)}=c>0,
 \end{equation} 
the nontrivial ICG density profile,
\begin{equation}\label{eq:129} 
\rho(\hat{r})_\mathrm{ICG}=\frac{4c}{\hat{T}}\left[1+2c\left(\frac{\hat{r}}{\hat{T}}\right)^2\right]^{-2},
\end{equation}
emerges.
It has an extremely fragile status between collapse and evaporation.
Indeed Abdalla and Rahimi Tabar \cite{AT98} had shown previously that the self-gravitating ICG in $\mathcal{D}=2$ undergoes a transition from a homogeneous phase to a collapsed phase at $\hat{T}_\mathrm{c}=\frac{1}{2}$ and that the (precarious) ICG state at $\hat{T}_\mathrm{c}$ has the density profile (\ref{eq:129}).
This nontrivial ICG density profile was also identified and used in other studies \cite{SC02, Aly94, AP99, TLPR10}.

The ICG profile (\ref{eq:129}) is relevant in the ILG context for $0<c\ll1$, where it can be identified as the solution of (\ref{eq:63}) for the case where $\hat{T}\to\hat{T}_\mathrm{c}$ from below.
This asymptotic solution also predicts the correct exponent value, $-2/\hat{T}\to-4$, in the power law (\ref{eq:82}).
Interestingly, the structure of (\ref{eq:129}) is a special case of an expression that emerges as quasi-equilibrium state from a kinetic model of self-gravitating systems in $\mathcal{D}=2,3$ \cite{TT10,Tash16}.

\begin{figure}[htb]
  \begin{center}
 \includegraphics[width=43mm]{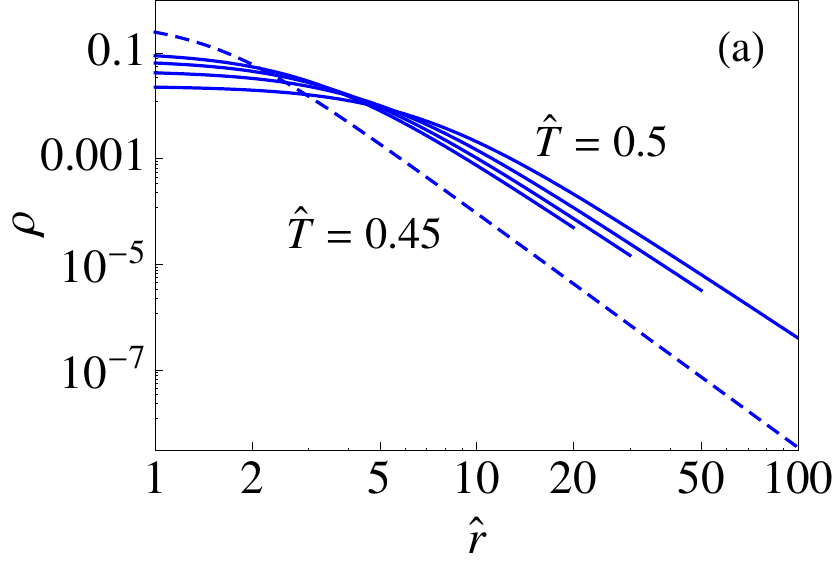}
 \includegraphics[width=43mm]{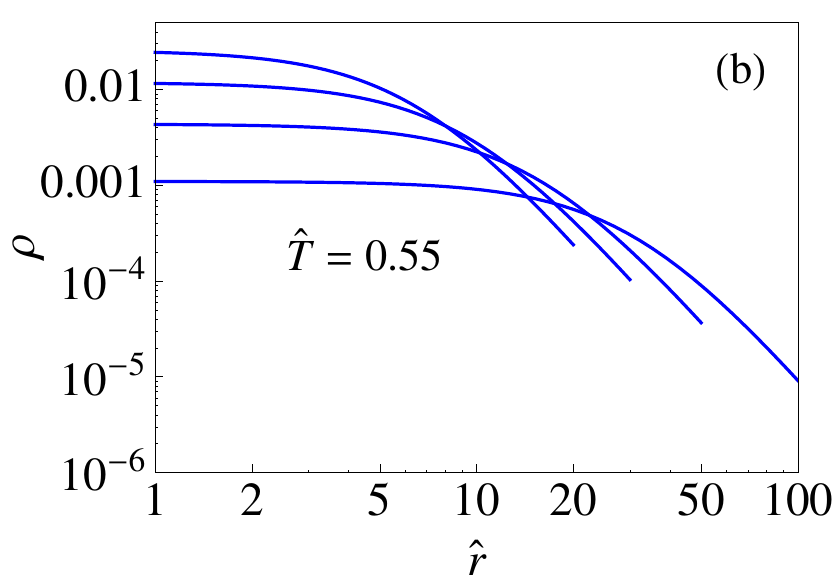}
\end{center}
\caption{Density profiles in $\mathcal{D}=2$ for the self-gravitating ILG confined to a disk-shaped space of radii $\hat{R}=20, 30, 50, 100$ at temperatures (a) $\hat{T}=0.45$ (dashed line), and $\hat{T}=\hat{T}_\mathrm{c}=0.5$ (solid lines), and (b) $\hat{T}=0.55$.}
  \label{fig:fig17}
\end{figure}

In Fig.~\ref{fig:fig17} we look at the stable and unstable self-gravitating ILG cluster from a different perspective.
We observe how, at constant $\hat{T}$, the density profile changes as we increase the radius $\hat{R}$ of the disk area to which the gas is being confined.
At $\hat{T}=0.45$, close below $\hat{T}_\mathrm{c}$, the cluster stays intact.
The profile change is imperceptibly small on the scale of the graph as the wall is moved from $\hat{R}=20$ to $\hat{R}=100$.
The power-law decay (\ref{eq:82}) is firmly established with near constant amplitude.

Performing the same isothermal expansion at $\hat{T}=\hat{T}_\mathrm{c}$ produces profiles that approach the shape of (\ref{eq:129}) with a gradually decreasing value of parameter $c$ and a power-law decay, $\sim\hat{r}^{-4}$, over a growing range of $\hat{r}$.
The evolution of the density profile under isothermal expansion is yet different at $\hat{T}=0.55$ close above $\hat{T}_\mathrm{c}$.
The asymptote (\ref{eq:82}) is no longer applicable.
The profile flattens out across a central area of increasing width and then curves downward near the confining wall.

The data in Fig.~\ref{fig:fig17} suggest that the ILG at fixed $1\ll\hat{R}<\infty$ and rising temperature undergoes a crossover centered at $\hat{T}_\mathrm{c}=\frac{1}{2}$ from a stable cluster with power-law profile (\ref{eq:82}) in the wings to a dilute gas with increasingly flat profile.
Only for $\hat{R}\to\infty$ does the crossover turn into the transition described previously.

Our ILG study shows that the hardcore repulsion does not affect the transition temperature.
The fact that close below $\hat{T}_\mathrm{c}$ the gas is already very dilute everywhere is consistent with that observation.
However, in strong contrast to the ICG, which suffers a gravitational collapse, the ILG exhibits a fluid phase at $\hat{T}<\hat{T}_\mathrm{c}$ with nontrivial density profile and $\hat{T}$-dependent power-law decay all the way down to $\hat{T}\to0$.

The self-gravitating FD gas, which shares with the ILG two key attributes, namely a strong short-range repulsion of sorts, relevant at high densities, and the ICG limit at low densities, exhibits similar phase behavior \cite{Chav02, Chav04} (see Appendix~\ref{sec:appe}).

\subsection{$\mathcal{D}=3$}\label{sec:Deq3}
Stable self-gravitating clusters at $\hat{T}>0$ of finite mass in $\mathcal{D}=3$ require confinement: $1<\hat{R}<\infty$.
The ILG and ICG both undergo transitions.
They are of a different nature than in $\mathcal{D}=2$.
We begin by examining the ILG. 
The results will alert us to the correct interpretation of the ICG data to be analyzed next.

The numerical analysis of (\ref{eq:63}) reveals that there are two parameter regimes.
In regime (i) for small $\hat{R}$, no precipitous events happen as $\hat{T}$ is lowered, but in regime (ii) for large $\hat{R}$ we find multiple solutions of (\ref{eq:63}) with identical conditions (\ref{eq:65}), (\ref{eq:67}).

\begin{figure}[t]
  \begin{center}
 \includegraphics[width=43mm]{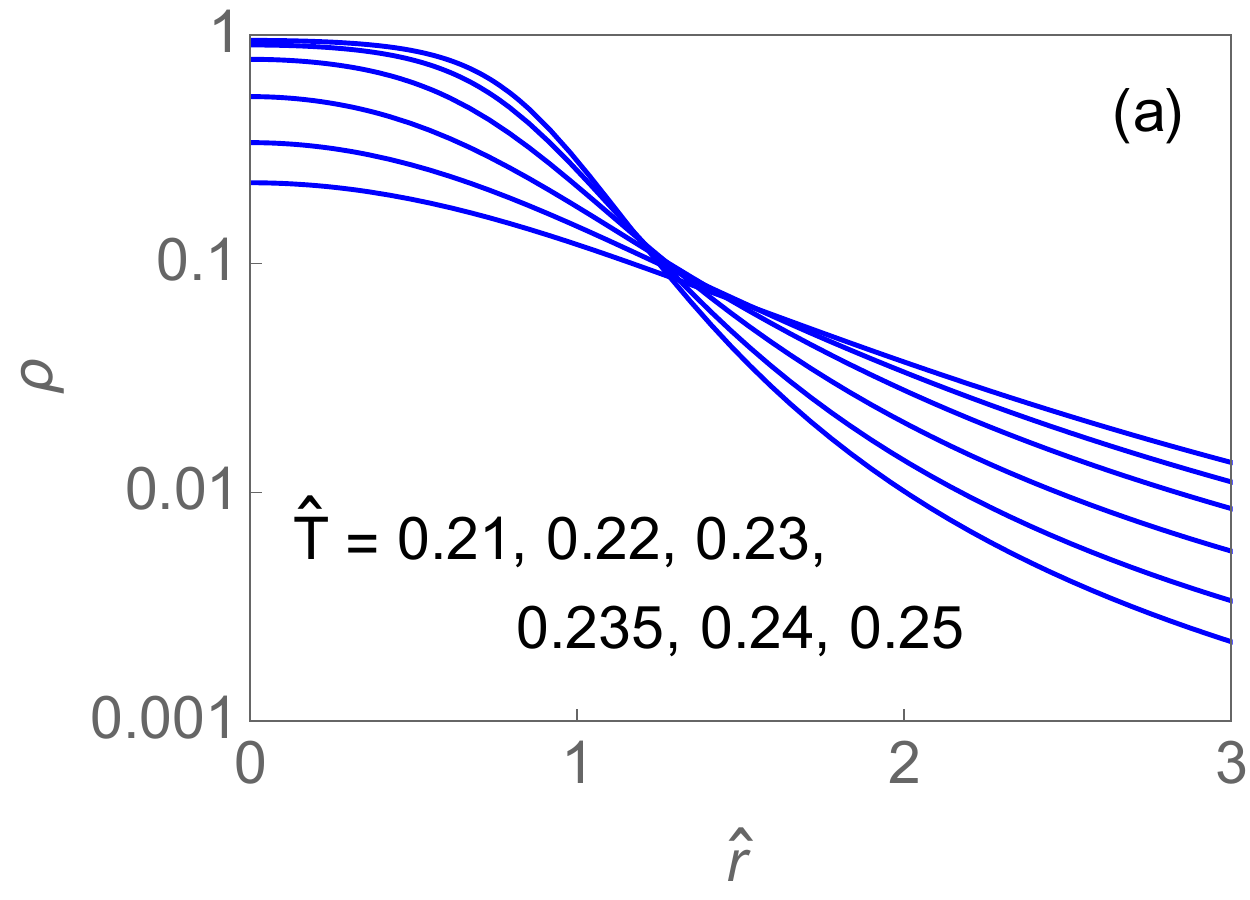}
 \includegraphics[width=43mm]{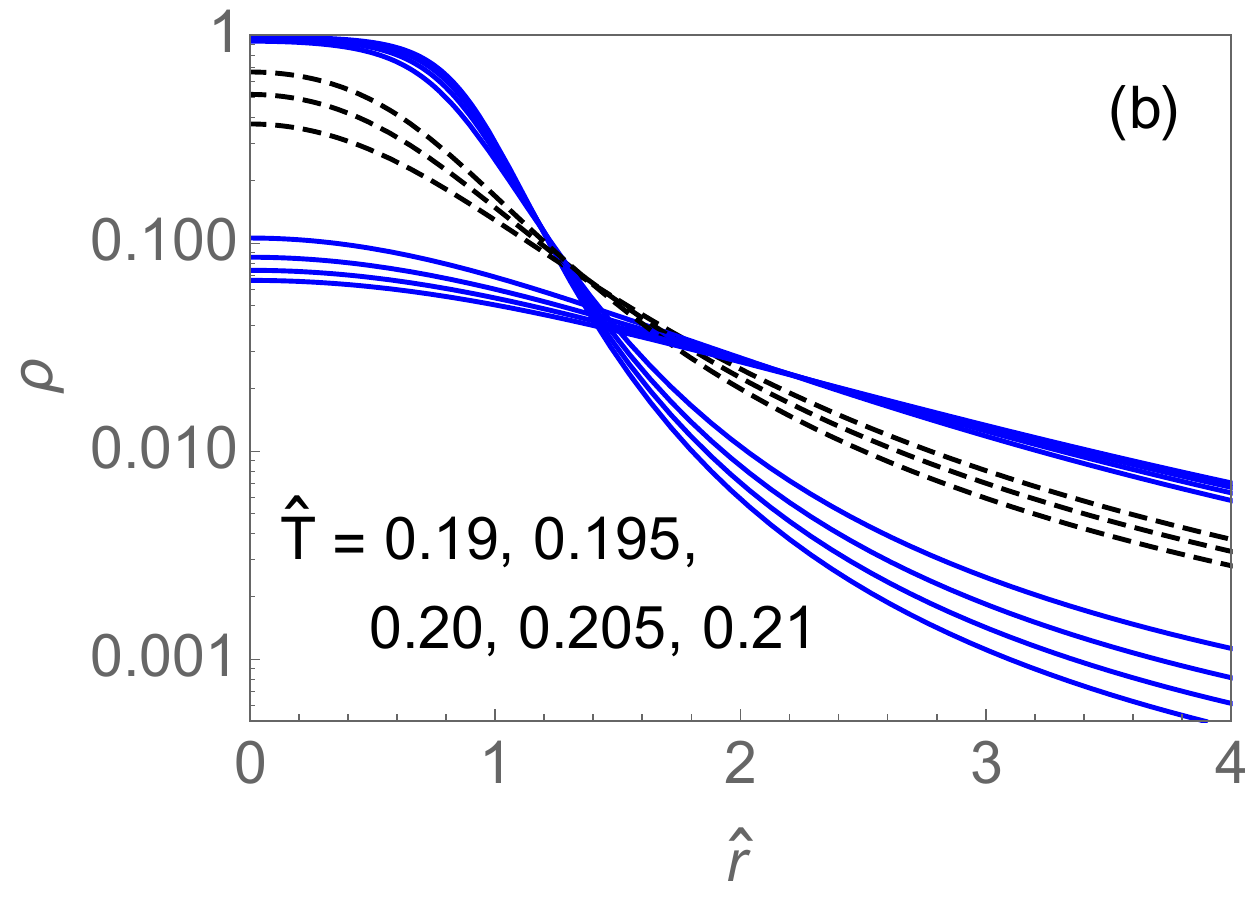}
\end{center}
\caption{Density profiles within the stated temperature range of the ILG confined to a spherical region of radius (a) $\hat{R}=3$ and (b) $\hat{R}=4$. Case (a) has a unique solution for all five values of $\hat{T}$ whereas case (b) has three solutions for the three intermediate values of $\hat{T}=0.195, 0.20, 0.205$. As $\hat{T}$ is being raised, $\rho(0)$ decreases (increases) for the solid (dashed) curves.}
  \label{fig:fig8}
\end{figure}

One case belonging to each regime is illustrated in Fig.~\ref{fig:fig8}.
When the ILG is confined to a sphere of (scaled) radius $\hat{R}=3$, we find a unique density profile as shown in panel (a).
We only show such profiles across a narrow range of $\hat{T}$.
Here their shape changes most rapidly with $\hat{T}$ while all changes remain gradual.
At the lower end of the interval, a cluster of near unit density with the hardcore repulsion visibly in action is present in outline.
This structure has all but disappeared at the upper end of the interval.
The maximum density (at $\hat{r}=0$) has dropped by a factor of five and the minimum density (at $\hat{r}=\hat{R}$) has increased by a similar factor.

In panel (b) we show how the density profile changes across a narrow interval of $\hat{T}$ for the same ILG confined to a somewhat larger sphere $(\hat{R}=4)$.
A unique density profile exists only outside this interval, namely at $\hat{T}\lesssim 0.19$ or $\hat{T}\gtrsim 0.21$.
In the high-$\hat{T}$ regime, the unique solution represents a relatively flat low-density gas profile $\rho_\mathrm{g}(\hat{r})$. 
That solution persists through the interval down to $\hat{T}\simeq 0.195$ and then disappears.
Likewise, in the low-$\hat{T}$ regime, a density profile $\rho_\mathrm{s}(\hat{r})$ describing a well formed cluster of close to unit density exists and continues to exist through the interval up to $\hat{T}\simeq 0.205$.
Both kinds of profiles are depicted as solid lines in Fig.~\ref{fig:fig8}(b).

For temperatures $0.195\lesssim\hat{T}\lesssim0.205$ the two aforementioned solutions coexist with a third solution of intermediate profile $\rho_\mathrm{i}(\hat{r})$ as shown dashed.
Of the three coexisting solutions at given $\hat{T}$, the equilibrium state is represented by the one with the lowest free energy.

We find that the lowest value of the free energy $\mathcal{F}(\hat{T})$ from (\ref{eq:111}) is assumed by either $\rho_\mathrm{s}(\hat{r})$ or $\rho_\mathrm{g}(\hat{r})$.
As we lower $\hat{T}$ across the interval of coexisting solutions, $\rho_\mathrm{g}(\hat{r})$ first has the lowest free energy.
Near the middle of that interval, the free energy of $\rho_\mathrm{s}(\hat{r})$ intersects that of $\rho_\mathrm{g}(\hat{r})$ and becomes the lowest.
At this temperature $\hat{T}_\mathrm{t}$, a first-order phase transition takes place. 
The free energy of $\rho_\mathrm{i}(\hat{r})$ has a higher value throughout the interval of coexisting profiles.

The coexisting solutions with increasing $\mathcal{F}(\hat{T})$ are stable, metastable, and unstable macrostates.
The $\hat{T}$-interval of coexisting solutions is bounded by spinodal points.
Here the metastable and unstable solutions coalesce and disappear.
Chavanis \cite{Chav14} (in his Fig.~7) showed three density profiles calculated in the framework of the microcanonical ensemble that correspond to three solutions such as shown in our Fig.~\ref{fig:fig8}(b) for the canonical ensemble.
The solution of intermediate central density in \cite{Chav14} is identified as being unstable, just as in our Fig.~\ref{fig:fig8}.

\begin{figure}[t]
  \begin{center}
 \includegraphics[width=43mm]{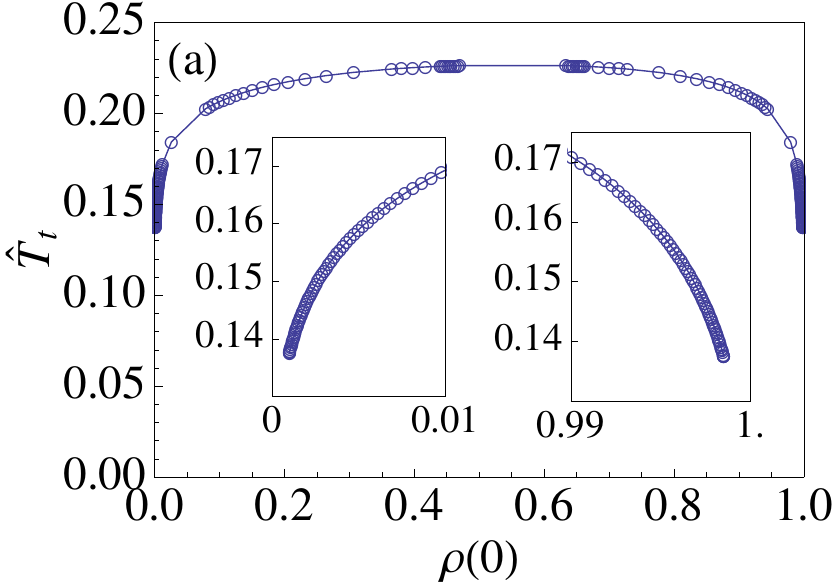}
 \includegraphics[width=41mm]{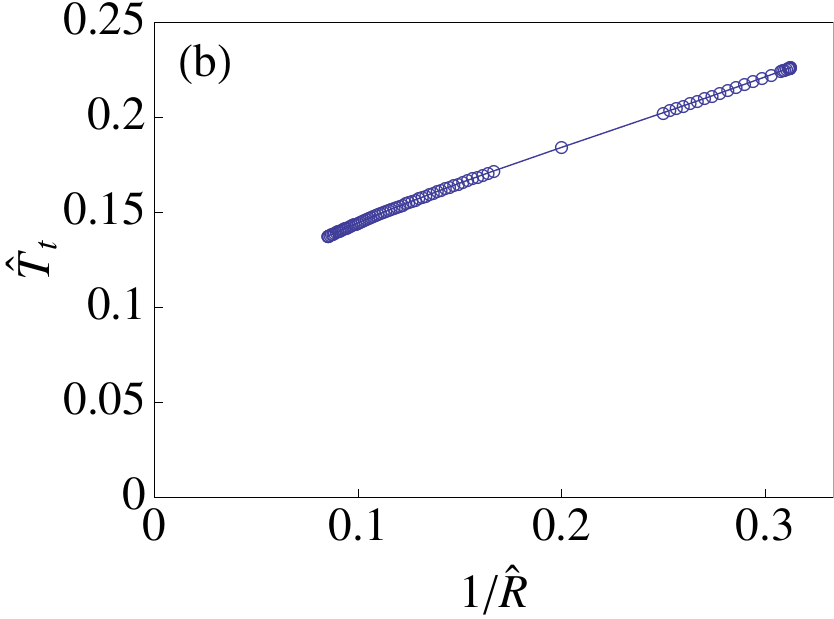}
\end{center}
\caption{(a) Guggenheim plot of densities $\rho_\mathrm{s}(0)$, $\rho_\mathrm{g}(0)$ for coexisting profiles at $\hat{T}_\mathrm{t}(\hat{R})$.
The insets highlight the approaches to the expected cusp singularities as $\hat{R}\to\infty$.
(b) Line of transition temperatures $\hat{T}_\mathrm{t}(\hat{R})$ versus inverse radius $\hat{R}^{-1}$ of confinement ending in a critical point $(\hat{R}^{-1}_\mathrm{c}, \hat{T}_\mathrm{c})$.
}
  \label{fig:fig21}
\end{figure}

The transition temperature $\hat{T}_\mathrm{t}$ increases with increasing $\hat{R}^{-1}$ as shown in Fig.~\ref{fig:fig21}(b).
The line of transition points terminates in a critical point $\hat{T}_\mathrm{c}\lesssim0.23$, pertaining to $\hat{R}_\mathrm{c}\gtrsim3.1$.
When we plot the values of $\rho_\mathrm{g}(0)$ and $\rho_\mathrm{s}(0)$ versus $\hat{T}_\mathrm{t}$ for $\hat{T}_\mathrm{t}\leq\hat{T}_\mathrm{c}$, a sort of Guggenheim plot emerges as shown in Fig.~\ref{fig:fig21}(a).

The line of data points in Fig.~\ref{fig:fig21}(b) is expected to bend down and reach  $\hat{T}_\mathrm{t}=0$ as $1/\hat{R}\to0$.
In that limit, the coexisting phases would be a solid of unit density and a fully sublimated gas of zero density.
The numerical analysis with sufficient precision becomes increasingly difficult as $\hat{R}$ gets larger.
Our data show the mere hint of the expected downward trend.

Now we turn to the ICG limit, which undergoes a gravitational collapse, just as its $\mathcal{D}=2$ counterpart does, but one of a different kind.
This phenomenon is well documented in previous work \cite{DdV07, Chav02a, Lali04, dVS06}.
A solution of (\ref{eq:132}) that is normalizable via (\ref{eq:133}) is found to exist only for temperatures $\bar{T}\doteq \hat{R}\hat{T}$ above the threshold value,  
\begin{equation}\label{eq:138} 
\bar{T}_\mathrm{C}=0.794422\ldots,
\end{equation}
implying, unlike in $\mathcal{D}=2$, that $\hat{T}_\mathrm{C}\to0$ as $\hat{R}\to\infty$.

\begin{figure}[htb]
  \begin{center}
 \includegraphics[width=70mm]{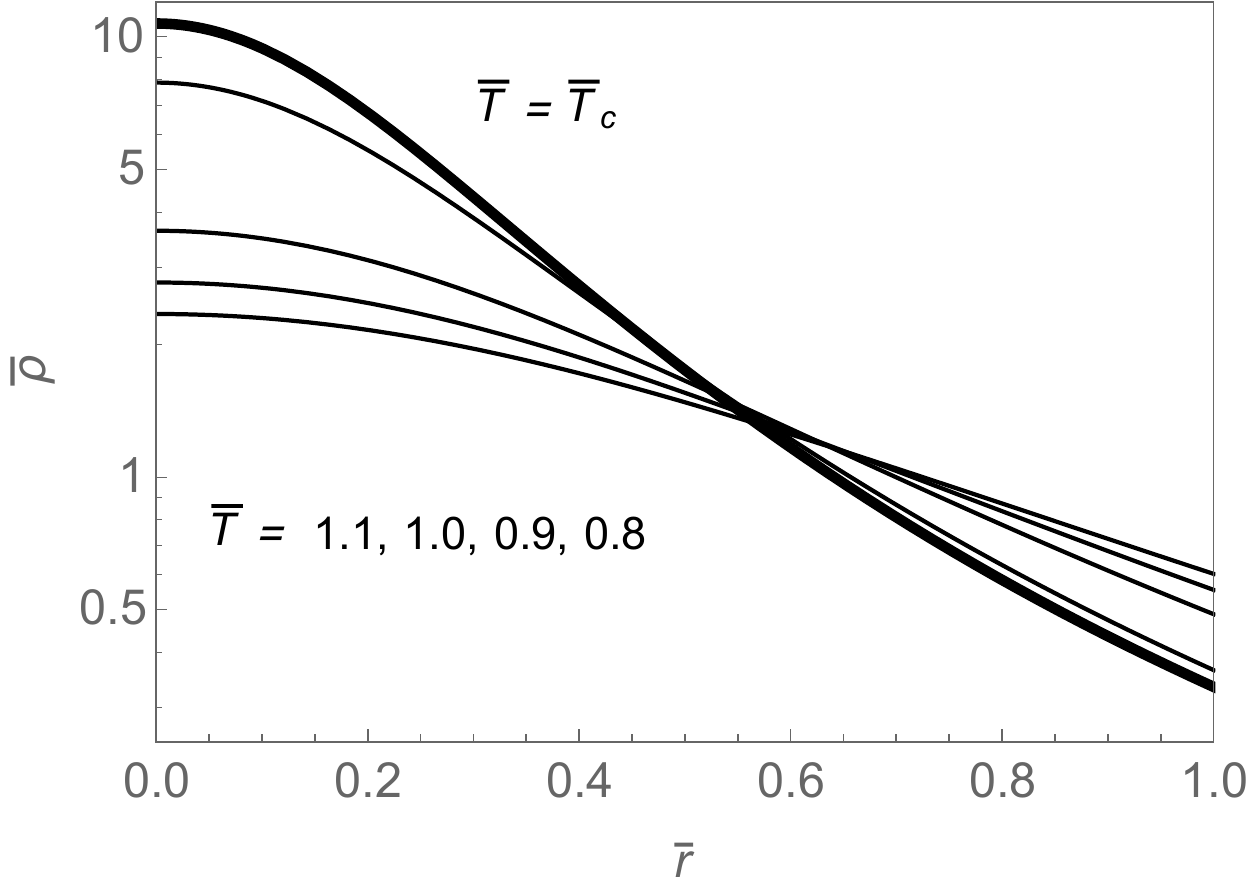}
\end{center}
\caption{Scaled density profile versus scaled radius of the confined ICG at $\bar{T}_\mathrm{C}$ (thick line), and four higher temperatures (thin lines). }
  \label{fig:fig9}
\end{figure}

The threshold ICG density profile is shown in Fig.~\ref{fig:fig9} along with profiles at selected $\bar{T}$ above $\bar{T}_\mathrm{C}$. 
As we lower $\bar{T}$ toward $\bar{T}_\mathrm{C}$ in steps of equal size a cluster appears to build up at an accelerated rate. 
However, in contrast to $\mathcal{D}=2$, that process does not come to its completion by gradually transforming the profile into a $\delta$-function. 
The collapse, which happens at $\bar{T}_\mathrm{C}$, is discontinuous, precipitated from a thermodynamic state that still pushes against the confining wall.

As in the ILG case discussed earlier, the existence limit at temperature (\ref{eq:138}) of a normalizable density profile out of (\ref{eq:132}) marks a sort of spinodal point for the gas phase rather than a transition point. 
This conclusion is indeed in line with the results of de Vega and Sanchez \cite{dVS06} based on different methodology. They predict two singular points,
\begin{subequations}\label{eq:110} 
\begin{align}\label{eq:110a} 
& \eta_\mathrm{C}\doteq\frac{Gm_\mathrm{c}^2N}{V^{1/3}k_\mathrm{B}T_\mathrm{C}} =1.561764, \\ \label{eq:110b}
& \eta_\mathrm{T}\doteq\frac{Gm_\mathrm{c}^2N}{V^{1/3}k_\mathrm{B}T_\mathrm{T}} =1.51024.
\end{align}
\end{subequations}
The former value, identified in \cite{dVS06} as the stability limit of the gas phase in the mean-field framework, matches our threshold temperature (\ref{eq:138}) to within 1ppm.
The latter value is identifed as the transition point to a collapsed state as indicated by singularities (e.g. in the isothermal compressibility) not captured by mean-field theory.

Unlike in the ILG case, here we lack the tool of comparing free energies for the purpose of identifying the transition temperature.
According to (\ref{eq:110b}) it is located at $\bar{T}_\mathrm{T}=0.8215\ldots$, some 3\% above $\bar{T}_\mathrm{C}$.
While the ILG transition is of first order, the ICG collapse features a discontinuity in free energy and might thus be classified as being of zeroth order \cite{dVS06}.

%
\section{Open systems}\label{sec:open syst}
%
We now examine the ILG and the ICG under conditions that characterize thermodynamically open systems, including systems with infinite mass.
We set $\hat{R}=\infty$ and use the scaled variables (\ref{eq:134}).
The ICG in $\mathcal{D}$ dimensions is then described by one universal density profile $\tilde{\rho}(\tilde{r})$, namely the solution of the Lane-Emden type ODE (\ref{eq:135}) with boundary conditions (\ref{eq:136}).
The ILG generalization,
\begin{equation}\label{eq:139} 
\frac{\tilde{\rho}''}{\tilde{\rho}}+\frac{\mathcal{D}-1}{\tilde{r}}\frac{\tilde{\rho}'}{\tilde{\rho}}
-\frac{1-2\rho_0\tilde{\rho}}{1-\rho_0\tilde{\rho}}\,\left(\frac{\tilde{\rho}'}{\tilde{\rho}}\right)^2
+\tilde{\rho}(1-\rho_0\tilde{\rho})=0,
\end{equation}
with $0<\rho_0<1$ and boundary conditions (\ref{eq:136}) again, describes a family of density profiles which includes the universal ICG profile as the limiting case $\rho_0\to0$.
Each solution reflects profiles across a range of temperatures.
The same profile may represent a stable, a metastable, or an unstable state at different temperatures.
The free energy $\mathcal{F}(\hat{T})$ from (\ref{eq:111}) does not produce a unique value for a given scaled profile. 
The parameter $\rho_0$, representing the density at the center of the cluster, is a substitute for the chemical potential, the commonly used control parameter for an open system.
Their relationship is explained in Sec.~\ref{sec:dif-equ}.

Our goal here is limited.
Describing how the main features of $\tilde{\rho}(\tilde{r})$ including the asymptotic decay depend on $\rho_0$ highlights the role of the hardcore repulsion in self-gravitating clusters.
Treating $\mathcal{D}$ as a continuous variable enables us to explore how the asymptotic decay crosses over between qualitatively different decay laws in $\mathcal{D}=1,2,3$.
The stability analysis of open systems is beyond the scope of this study.
It would require that we unfold the scaled profiles and develop additional tools.

\subsection{$\mathcal{D}=1$}\label{sec:Deq1op}
\begin{figure}[htb]
  \begin{center}
 \includegraphics[width=41.5mm]{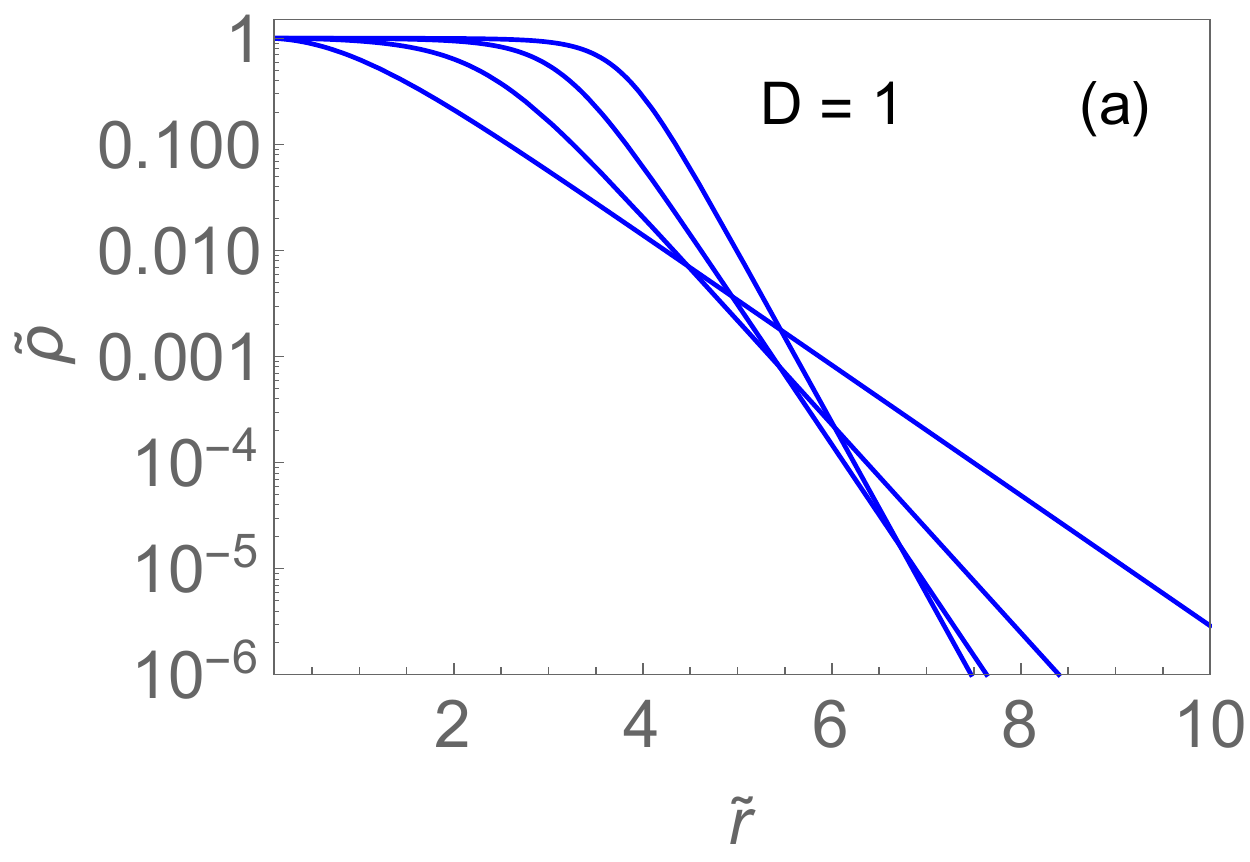}
  \includegraphics[width=42.5mm]{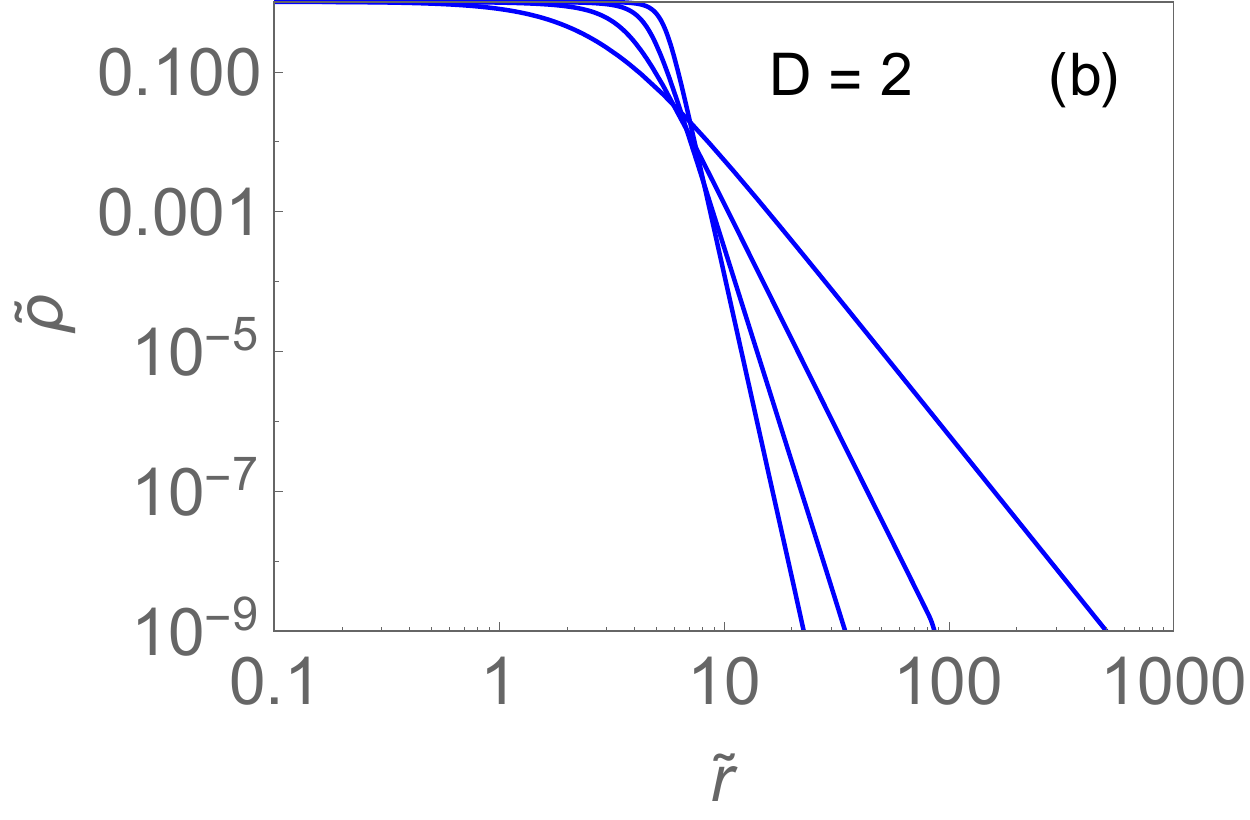}
   
 \includegraphics[width=42mm]{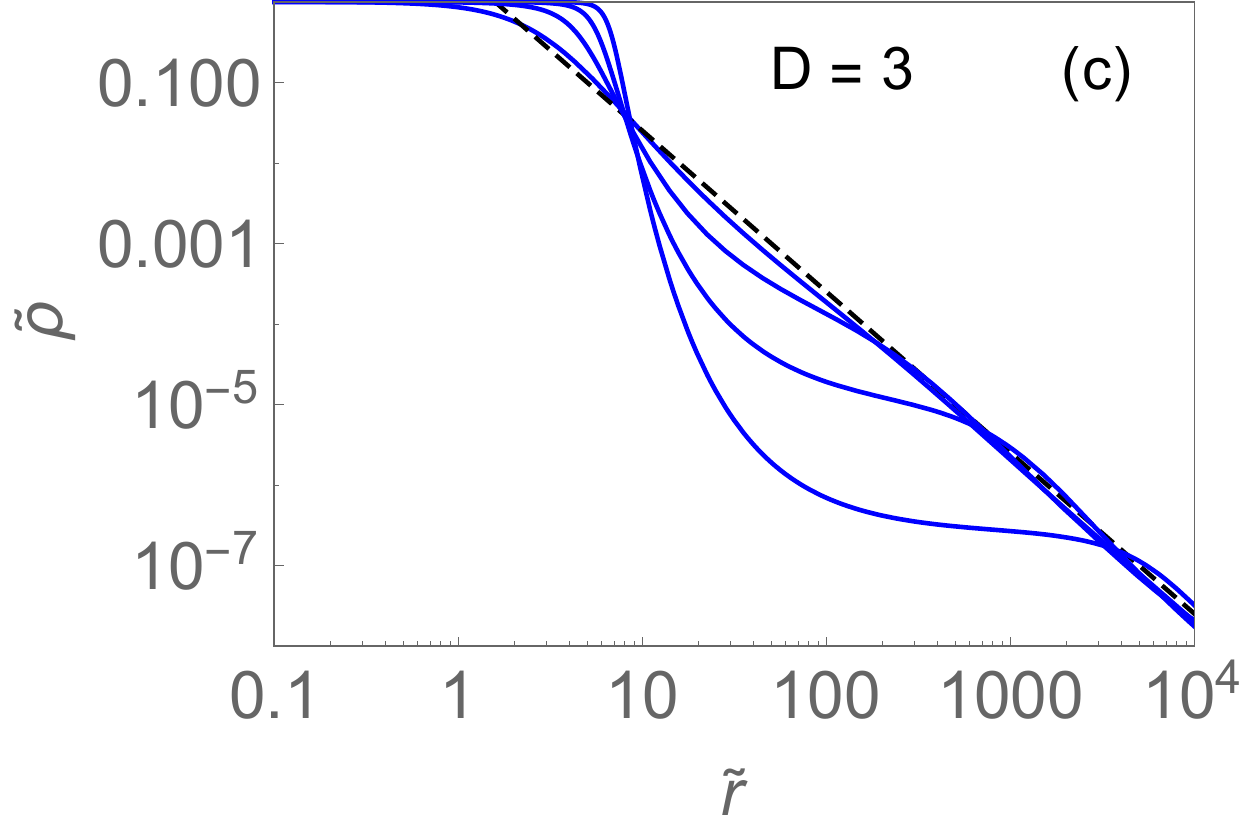}
\end{center}
\caption{One-parameter family of density profiles for an open ILG gas in (a) $\mathcal{D}=1$, (b) $\mathcal{D}=2$, and (c) $\mathcal{D}=3$.
The parameter values are $\rho_0=0, 0.9, 0.99, 0.999$ in each case. 
The limiting ICG profile, $\rho_0\to0$, is the one decaying most slowly in $\mathcal{D}=1,2$ and the one reaching asymptotic behavior first in $\mathcal{D}=3$.
The dashed line in (c) represents (\ref{eq:143}).}
  \label{fig:fig20}
\end{figure}

Open ILG clusters in $\mathcal{D}=1$ are represented by a one-parameter family of solutions of (\ref{eq:139}). 
The density decays exponentially as illustrated in Fig.~\ref{fig:fig20}(a).
The limiting case $\rho_0\to0$ represents the universal ICG profile:
\begin{equation}\label{eq:140} 
\tilde{\rho}(\tilde{r})=\mathrm{sech}^2\big(\tilde{r}/\sqrt{2}\big),
\end{equation}
the analytic solution of (\ref{eq:135}).
This universal profile covers dilute clusters of different (average) sizes at different temperatures by virtue of the scaling (\ref{eq:134}).

With $\rho_0$ growing from zero the decay rate increases monotonically. 
A solid-like core emerges gradually and grows in size while the surrounding atmosphere thins out more and more quickly with distance from the solid core.
All solutions of (\ref{eq:139}) for $\mathcal{D}=1$ describe clusters of finite (average) mass.
In Appendix~\ref{sec:appa} we prove that the decay law must be of the form
\begin{equation}\label{eq:145} 
\tilde{\rho}(\tilde{r})_\mathrm{as} \sim  e^{-\nu_1(\rho_0)\tilde{r}}.
\end{equation}
In Appendix~\ref{sec:appb} we derive the analytic solution of (\ref{eq:139}).
It is most concisely expressed via the inverse function,
\begin{equation}\label{eq:109} 
\tilde{r}(\tilde{\rho})=\sqrt{\frac{\rho_0}{2}}\int_{\tilde{\rho}}^{1}\frac{d\tilde{\rho}'}{\tilde{\rho}'(1-\rho_0\tilde{\rho}')\sqrt{\ln\frac{1-\rho_0\tilde{\rho}'}{1-\rho_0}}}~ 
: 0\leq\tilde{\rho}\leq 1.
\end{equation}
The exponential decay rate extracted from (\ref{eq:109}),
\begin{equation}\label{eq:108} 
\nu_1(\rho_0)=\sqrt{-\frac{2}{\rho_0}\ln(1-\rho_0)},
\end{equation}
is consistent with (\ref{eq:140}) in the limit $\rho_0\to0$.

\subsection{$\mathcal{D}=2$}\label{sec:Deq2op}
Several profiles for open ILG clusters in $\mathcal{D}=2$ are shown in Fig.~\ref{fig:fig20}(b) including the limiting ICG case, ${\rho_0\to0}$.
The analytic solution of (\ref{eq:135}) reads
\begin{equation}\label{eq:141} 
\tilde{\rho}(\tilde{r})=\big(1+\tilde{r}^2/8\big)^{-2}.
\end{equation}
The ILG power-law decay is rigorously established in Appendix~\ref{sec:appa},
\begin{equation}\label{eq:142}
\tilde{\rho}(\tilde{r})_\mathrm{as}\sim\tilde{r}^{-\nu_2(\rho_0)},
\end{equation}
but the function $\nu_2(\rho_0)$ is not exactly known.

\begin{figure}[b]
  \begin{center}
\includegraphics[width=60mm]{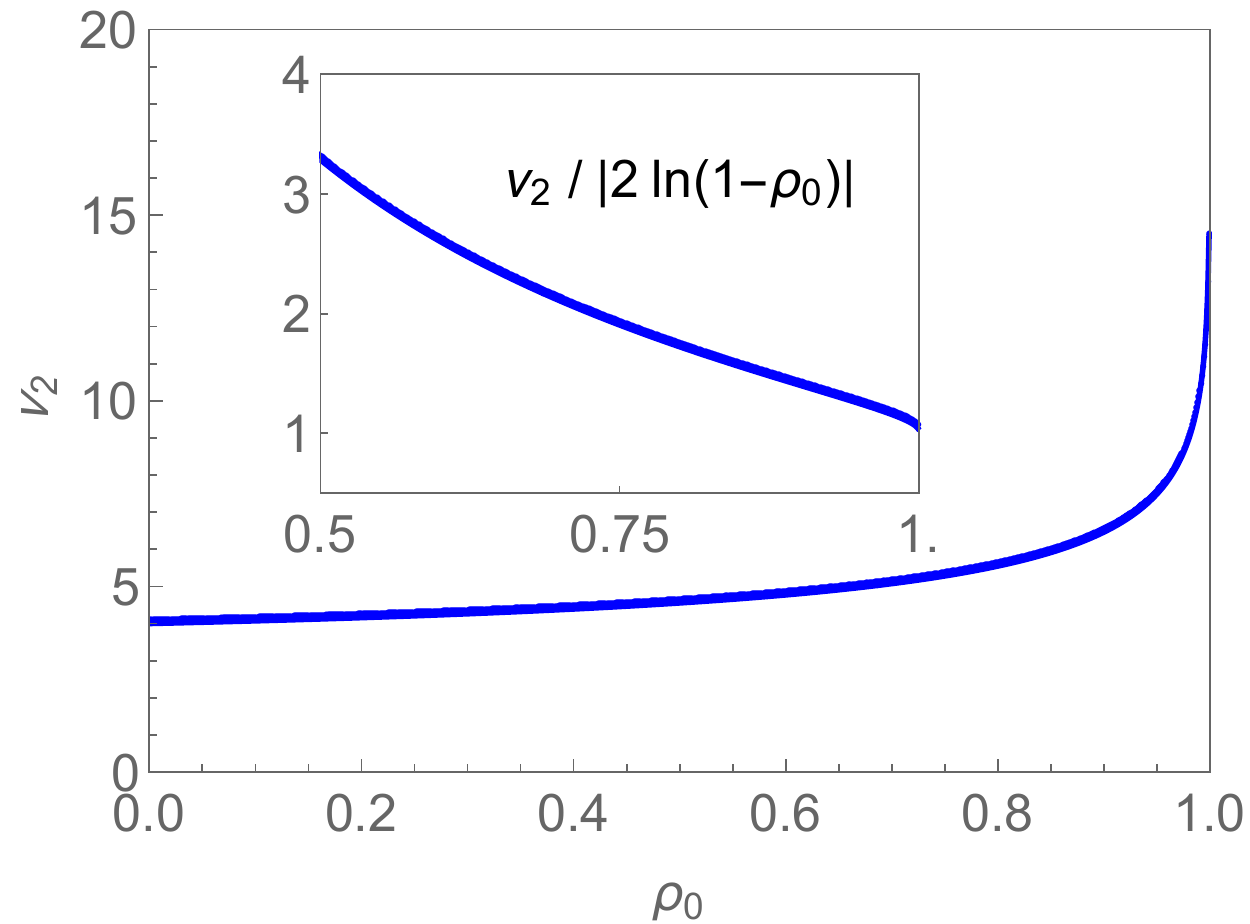}
\end{center}
\caption{Data for the exponent $\nu_2(\rho_0)$ as inferred from (\ref{eq:a6}) for open clusters in $\mathcal{D}=2$.}
  \label{fig:fig22}
\end{figure}

The data in Fig.~\ref{fig:fig22} connect with the known ICG limit, $\nu_2(0)=4$, and strongly suggest a monotonic increase with a weak divergence at $\rho_0=1$.
By an iteration process as described in Ref.~\cite{TA15} we can prove that
\begin{equation}\label{eq:107} 
 \nu_2(\rho_0) \stackrel{\rho_0\to1}{\longrightarrow} -2\ln(1-\rho_0),
\end{equation}
which is consistent with the data as displayed in the inset, actually holds rigorously.
This power-law decay guarantees that all clusters thus described have a finite mass.  

\subsection{$\mathcal{D}=3$}\label{sec:Deq3op}
In Sec.~\ref{sec:Deq3} we have discussed the contrasting behavior of ILG and ICG clusters under confinement.
Here we examine solutions of (\ref{eq:139}) in $\mathcal{D}=3$ representing ILG clusters of infinite mass. 
Profiles with a wide range of parameter values $\rho_0$ are shown in Fig.~\ref{fig:fig20}(c).

The limiting ICG universal curve for $\rho_0\to0$ is the profile of the well known Bonnor-Ebert sphere \cite{Chand42} and shows the characteristic asymptotic power-law decay, 
\begin{equation}\label{eq:143} 
\tilde{\rho}(\tilde{r})\sim 2\tilde{r}^{-2}.
\end{equation}
This decay law also holds for the ILG with any $\rho_0<1$ as proven in Appendix~\ref{sec:appa}.

The effects of the hardcore repulsion in the ILG profiles are quite intriguing.
With increasing $\rho_0$ we see the gradual emergence of a structure with three layers: a solid-like core surrounded by a shell of dilute atmosphere with slowly varying density out to some well-defined radius, where it crosses over into the halo characteristic of the Bonnor-Ebert asymptotic profile (\ref{eq:143}).
Somewhat similar density profiles have previously been calculated for the FD gas in $\mathcal{D}=3$ \cite{IR88}.

\subsection{Asymptotics for varying $\mathcal{D}$}\label{sec:Dvar}
When we consider the ODE (\ref{eq:135}) with boundary conditions (\ref{eq:136}) representing the scaled density profile of an open ICG cluster we are left with a single parameter $\mathcal{D}$ that can be varied continuously, touching on the three integer values $\mathcal{D}=1,2,3$ for which physical realizations exist or, at least, are conceivable.
The asymptotic decay of the scaled density is qualitatively different for these three landmarks as noted before:
\begin{equation}\label{eq:146} 
\tilde{\rho}(\tilde{r})\sim \left\{ \begin{array}{ll}
e^{-\sqrt{2}\,\tilde{r}} &:~ \mathcal{D}=1, \rule[-2mm]{0mm}{5mm}\\
\tilde{r}^{-4} &:~ \mathcal{D}=2,\rule[-2mm]{0mm}{5mm} \\
\tilde{r}^{-2} &:~ \mathcal{D}=3.\rule[-2mm]{0mm}{5mm}
\end{array} \right.
\end{equation}
How does the asymptotic decay law, which, as shown in Sec.~\ref{sec:Asym}, also holds for ILG clusters, vary between and beyond these integer dimensions? 
As it turns out, we again find three qualitatively different answers.

\begin{figure}[htb]
  \begin{center}
 \includegraphics[width=41.5mm]{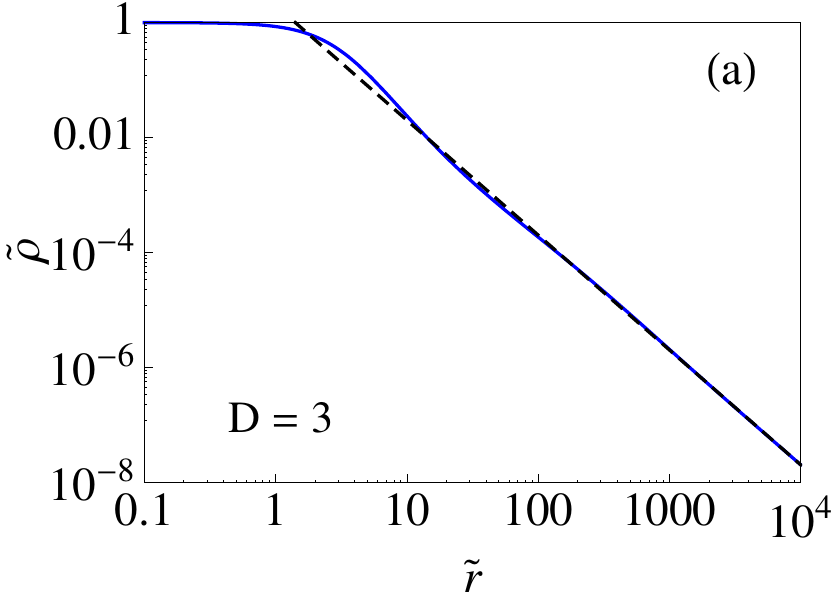}
 \includegraphics[width=42.5mm]{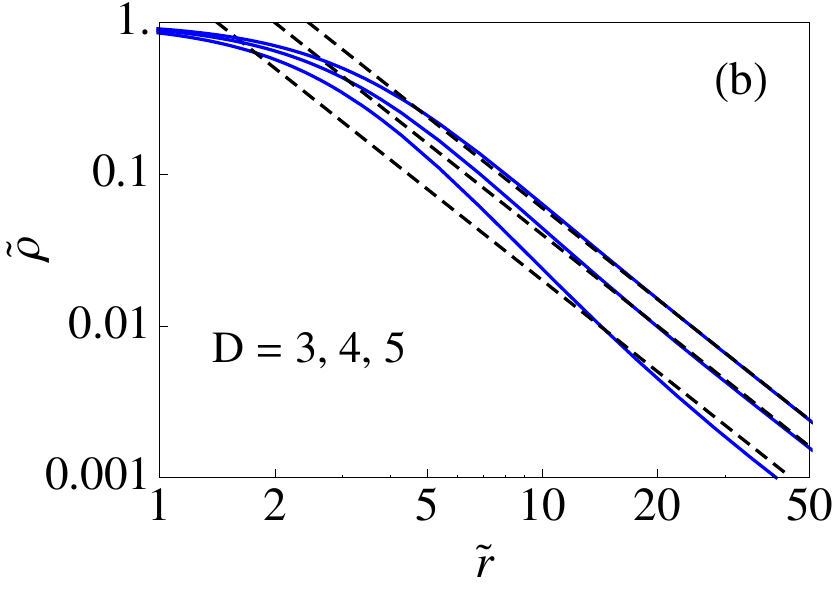}
\end{center}
\caption{Density profiles of an open ICG cluster (a) in $\mathcal{D}=3$ over a long range of radius and (b) in $\mathcal{D}=3,4,5$ over a shorter range. The dashed lines represent the asymptote (\ref{eq:147}). }
  \label{fig:fig23}
\end{figure}

We begin by exploring the range $\mathcal{D}\geq3$.
It is straightforward to show that the ansatz,
\begin{equation}\label{eq:147} 
\tilde{\rho}(\tilde{r})\sim a\tilde{r}^{-\alpha},
\end{equation}
is an asymptotic solution of (\ref{eq:135}) if we set
\begin{equation}\label{eq:148} 
\alpha=2,\quad a=2(\mathcal{D}-2),
\end{equation}
implying that the inverse-square decay law remains intact albeit with a change in meaning.
Successive shells of equal width contain the same amount of dilute gas in $\mathcal{D}=3$ whereas that amount increases with $\tilde{r}$ in $\mathcal{D}>3$.
We also observe (in Fig.~\ref{fig:fig23}) that the asymptotic decay (\ref{eq:148}) sets in earlier as the dimensionality increases from $\mathcal{D}=3$. 
The mild deviations from the asymptotic decay, most conspicuous in $\mathcal{D}=3$, are reminiscent of damped oscillations. 

The asymptotic decay (\ref{eq:147}) with (\ref{eq:148}) remains valid also for $2<\mathcal{D}<3$ but here the deviations are of a different nature.
What makes $\mathcal{D}=3$ a landmark dimensionality is that the relative importance of the second and third terms in (\ref{eq:135}) switches.
We have $\tilde{\rho}'/\tilde{\rho}=2/\tilde{r}$, which is to be compared with $(\mathcal{D}-1)/\tilde{r}$.

\begin{figure}[htb]
  \begin{center}
 \includegraphics[width=42mm]{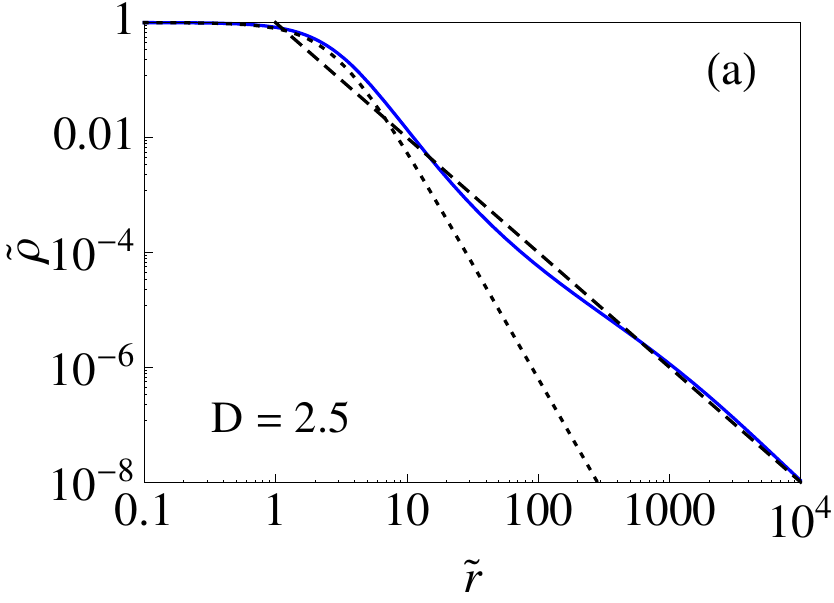}
 \includegraphics[width=42mm]{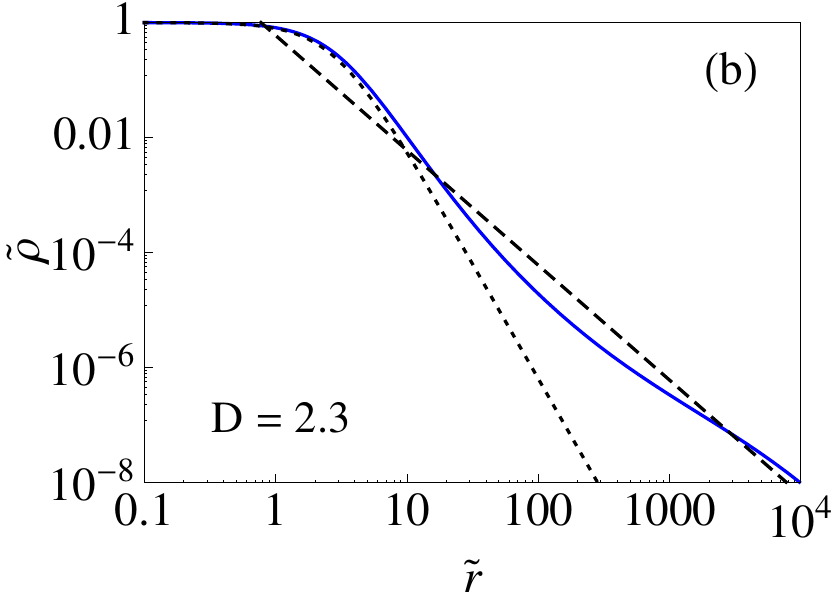}

\includegraphics[width=42mm]{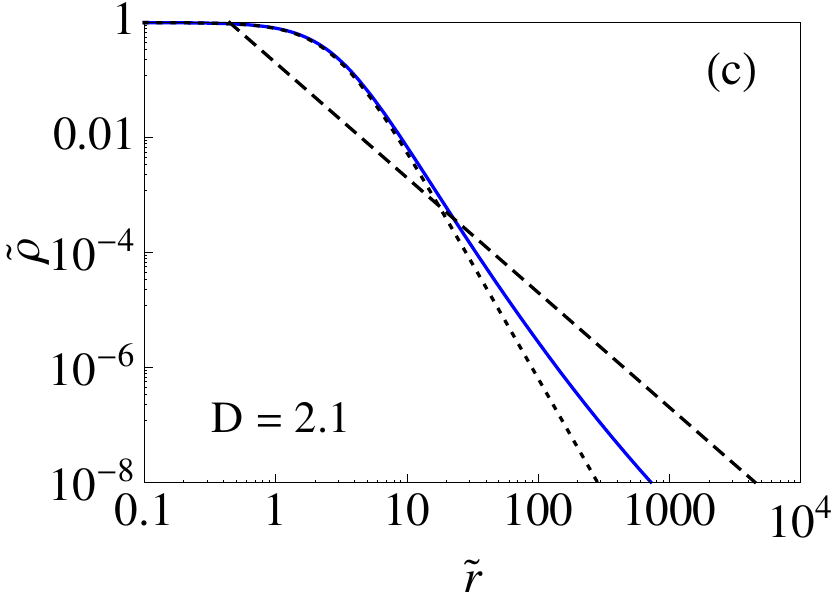}
\includegraphics[width=42mm]{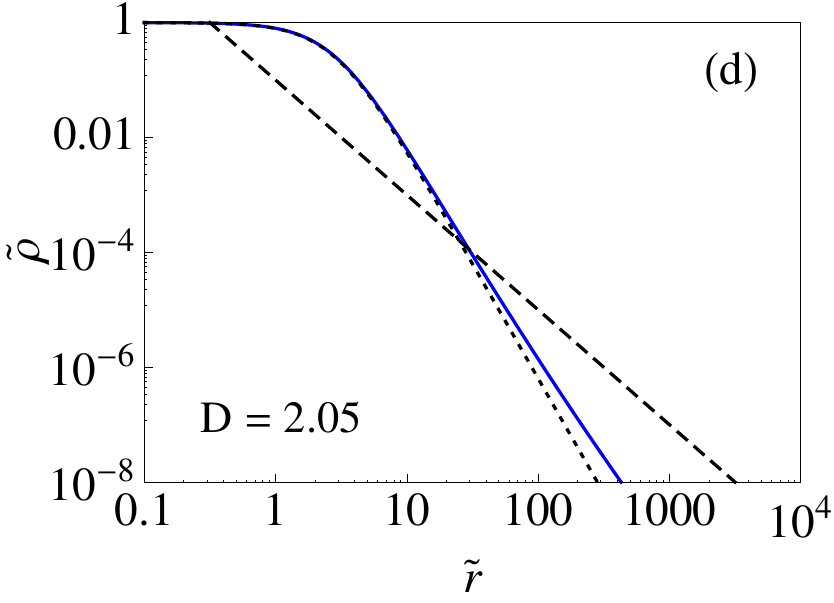}
\end{center}
\caption{Crossover between incipient asymptotics, $\sim\tilde{r}^{-4}$, at small to intermediate $\tilde{r}$ and true asymptotics, $\sim \tilde{r}^{-2}$, at large $\tilde{r}$ of an open ICG cluster in $2<\mathcal{D}<3$. The dotted line represents the exact result (\ref{eq:141}) in $\mathcal{D}=2$ and the dashed line the exact asymptote (\ref{eq:143}) in $\mathcal{D}=3$.}
  \label{fig:fig24}
\end{figure}

The interpolation between the two distinct power laws of (\ref{eq:146}) is not realized by a variable exponent but by a crossover between the faster power-law decay at small and intermediate radii and the slower power-law decay at larger radii.
This is illustrated in Fig.~\ref{fig:fig24}.
As $\mathcal{D}$ decreases the crossover radius grows and reaches infinity for $\mathcal{D}=2$.

\begin{figure}[htb]
  \begin{center}
 \includegraphics[width=43mm]{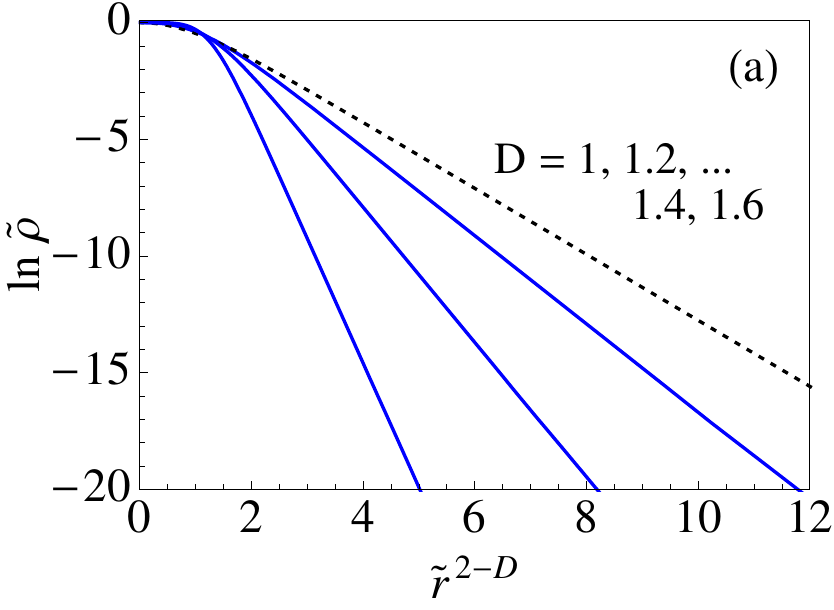}
 \includegraphics[width=41mm]{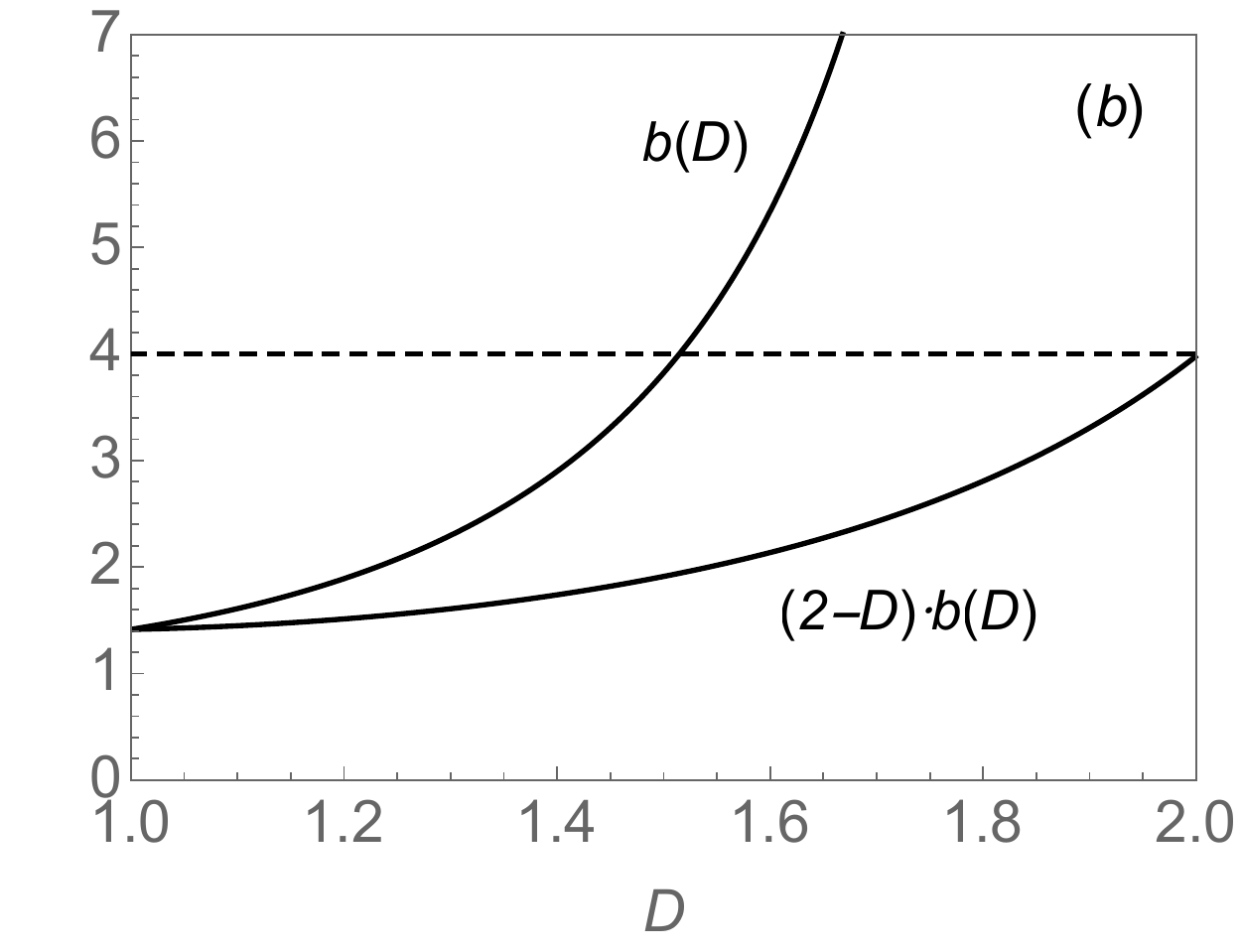}
\end{center}
\caption{Stretched exponential asymptotics (\ref{eq:149}) of an open ICG cluster in $1<\mathcal{D}<2$. The dotted line represents the exact result (\ref{eq:140}) in $\mathcal{D}=1$. The data in panel (b) are extracted from the asymptotes of curves such as shown in panel (a).}
  \label{fig:fig25}
\end{figure}

The interpolation between exponential decay and power-law decay in the range ${1<\mathcal{D}<2}$ is yet of a different kind.
In this regime our numerical analysis of (\ref{eq:135}) points to a stretched exponential decay,
\begin{equation}\label{eq:149} 
\tilde{\rho}(\tilde{r})\sim \exp\left(-b(\mathcal{D})\,\tilde{r}^{\beta(\mathcal{D})}\right),
\end{equation}
with
\begin{subequations}\label{eq:105} 
\begin{align}\label{eq:105a}
& \beta(\mathcal{D})=2-\mathcal{D},\\ \label{eq:105b}
& \lim_{\mathcal{D}\to1}b(\mathcal{D})=\sqrt{2}, \\ \label{eq:105c}
& \lim_{\mathcal{D}\to2}\beta(\mathcal{D})b(\mathcal{D})=4,
\end{align}
\end{subequations}
as illustrated in Fig.~\ref{fig:fig25}.
The results (\ref{eq:105a}) and (\ref{eq:105b}) are rigorous.
The data in Fig.~\ref{fig:fig25}(b) strongly suggest that (\ref{eq:105c}) is accurate.
The case $\mathcal{D}=2$ is the most delicate for this type of analysis.
It also represents the transition from clusters with finite average mass to infinite mass.

%
\section{Summary and outlook}\label{sec:sum-out}
%
In this work we have been advocating the hitherto neglected case of the lattice gas as a useful model for the study of density profiles in self-gravitating material clusters of dimensionality $\mathcal{D}=1,2,3$ at thermal and mechanical equilibrium.
The ILG equation of state (\ref{eq:117}) has a simple structure, includes the ICG of classical point particles as a limiting case, and prevents the (artificial) gravitational collapse of point particles by a robust hardcore repulsive force.

The dual (necessary) conditions of mechanical and thermal equilibrium have led to a second-order ODE for the density profile with several parameters.
In closed systems the ODE has the form (\ref{eq:63}) and in open systems the form (\ref{eq:139}).
One parameter is the dimensionality of the space, with discrete values $\mathcal{D}=1,2,3$ in most of the work, and treated as a continuous parameter in Sec.~\ref{sec:Dvar}.
A second parameter is the temperature.
For open systems, a third parameter is the chemical potential, expressed via the density at the center of the cluster.
For closed systems with wall confinement, the radius of the available space  is a third parameter. 

Sufficient conditions for thermal equilibrium require, in the framework of our study, an expression of free energy as a discriminant for multiple solutions of (\ref{eq:63}). 
One contribution to that free energy is the gravitational self-energy, for which we have derived an expression in the form of a density functional that works for the ILG  in all dimensions and is equivalent to the commonly used expression in $\mathcal{D}=3$.

We have calculated some exact results for density profiles of the ILG, supplemented by graphical results of numerical integrations.
In most cases we have been able to derive the long-distance asymptotic decay of density profiles exactly. 
We have also identified a continuous transition in the unconfined ILG for $\mathcal{D}=2$ and a discontinuous transition in the confined ILG for $\mathcal{D}=3$.

Multiple contacts with the ICG, which emerges from the ILG in the low-density limit, and with models that employ alternative short-distance regularizations, have been established in Secs.~\ref{sec:equi-state}, \ref{sec:open syst} and, especially, in Appendix \ref{sec:appe}, confirming a host of results from previous studies. 

Our focus on density profiles, supplemented by some profiles of pressure and potential, will be kept in an extension of this work that examines rotating ILG clusters. 
The competing gravitational and centrifugal forces produce a plethora of new phenomena that have scarcely been investigated, particularly in low dimensions \cite{roclus}. 

\appendix

%
\section{Gravitational self-energy}\label{sec:appc}
\begin{figure}[htb]
  \begin{center}
 \includegraphics[width=70mm]{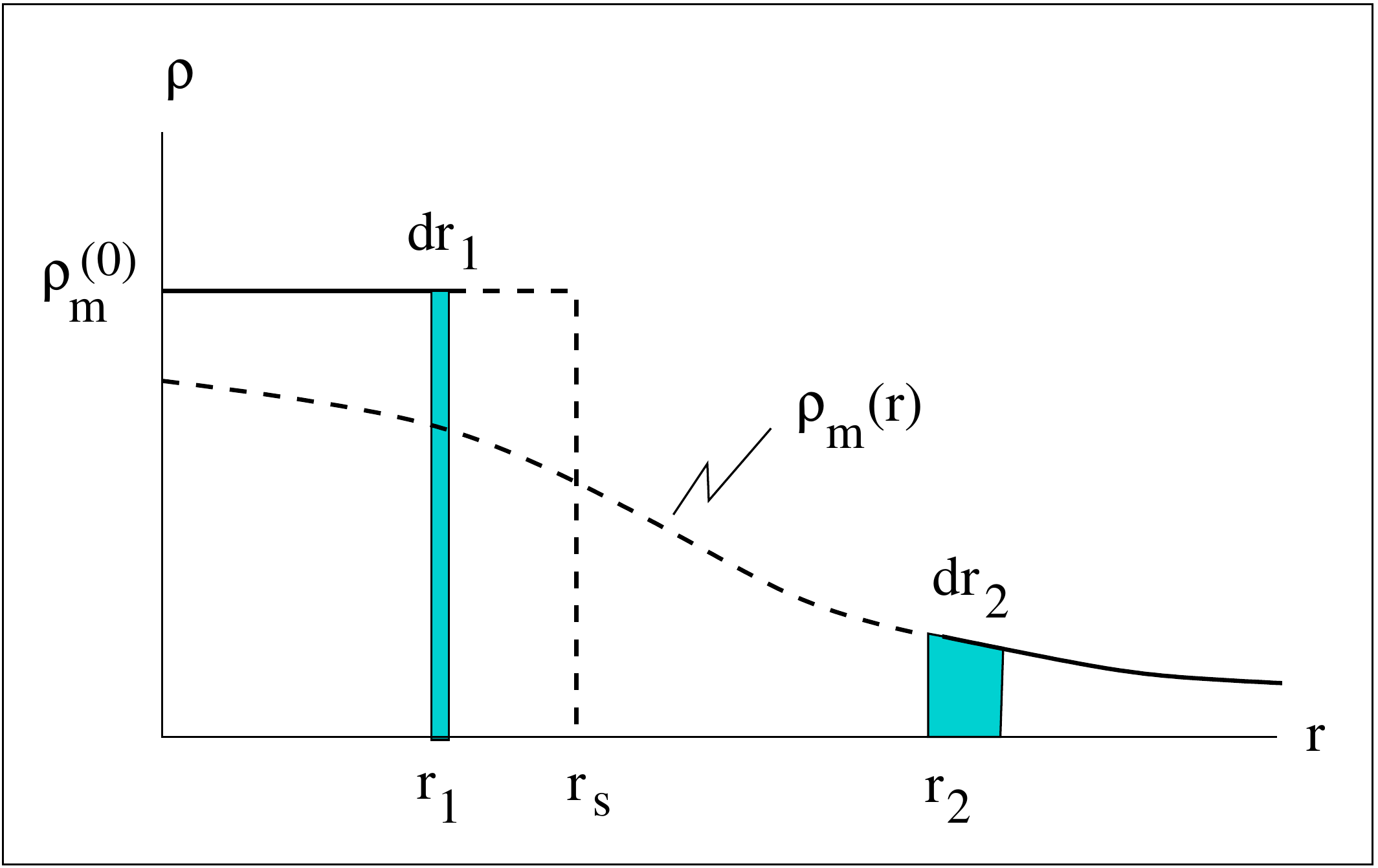}
\end{center}
\caption{Change in gravitational self-energy $dU_\mathrm{S}$ calculated as work performed against gravity when a thin layer of mass $dm$ is being translocated from radius $r_1$ to radius $r_2$.}
  \label{fig:fig27}
\end{figure}

%
The gravitational self-energy $U_\mathrm{S}$ relative to its value in the ground state of a symmetric ILG cluster with finite mass is the first term in the Helmholtz free energy (\ref{eq:111}).
We construct $U_\mathrm{S}$ as the work performed against gravity when mass of maximum density $\rho_\mathrm{m}^{(0)}$ is moved in the shape of thin layers 
from position $r_1$ in the ground-state profile to position $r_2$ in any given mass-density profile $\rho_\mathrm{m}(r)$.
This process of disassembling the reference profile and reassembling a generic profile is illustrated in Fig.~\ref{fig:fig27}.
For clarity we do the scaling at the end.

The increment of self-energy is
\begin{equation}\label{eq:c1} 
dU_\mathrm{S}=dm\big[\mathcal{U}(r_2)-\mathcal{U}(r_1)\big],
\end{equation}
where 
\begin{equation}\label{eq:c2} 
dm=\mathcal{A_D}r_1^{\mathcal{D}-1}dr_1\rho_\mathrm{m}^{(0)}=
 \mathcal{A_D}r_2^{\mathcal{D}-1}dr_2\rho_\mathrm{m}(r_2).
\end{equation}
The potential at either position depends on the solid mass,
\begin{equation}\label{eq:c16} 
 m_1=\int_0^{r_1}dr\mathcal{A_D}r^{\mathcal{D}-1}\rho_\mathrm{m}^{(0)}
 =\frac{\mathcal{A_D}}{\mathcal{D}}r_1^\mathcal{D}\rho_\mathrm{m}^{(0)},
\end{equation}
at $r\leq r_1$ only:
\begin{align}\label{eq:c3} 
\mathcal{U}(r_2)-\mathcal{U}(r_1)
&=Gm_1\int_{r_1}^{r_2}\frac{dr}{r^{\mathcal{D}-1}} \nonumber \\
&\hspace{-10mm}=\frac{G}{\mathcal{D}-2}\left(\frac{\mathcal{A_D}}{\mathcal{D}}
\rho_\mathrm{m}^{(0)}r_1^\mathcal{D}\right)\Big(r_1^{2-\mathcal{D}}
-r_2^{2-\mathcal{D}}\Big).
\end{align}
Mass conservation as reflected in (\ref{eq:c2}) expresses $r_1$ as a function of $r_2$:
\begin{equation}\label{eq:c4} 
r_1^\mathcal{D}=\mathcal{D}\int_0^{r_2}dr\,r^{\mathcal{D}-1}\rho(r),\quad
\rho(r)\doteq\frac{\rho_\mathrm{m}(r)}{\rho_\mathrm{m}^{(0)}},
\end{equation}
with the latter in the role of integration variable for (\ref{eq:c1}):
\begin{equation}\label{eq:c5} 
U_\mathrm{S}=\frac{\mathcal{D}Gm_\mathrm{t}^2}{\mathcal{D}-2}
r_\mathrm{s}^{-2\mathcal{D}}\int_0^\infty dr_2\,\rho(r_2)
\Big[r_1^2r_2^{\mathcal{D}-1}-r_1^\mathcal{D}r_2\Big],
\end{equation}
where $m_\mathrm{t}=(\mathcal{A_D/D})r_\mathrm{s}^\mathcal{D}\rho_\mathrm{m}^{(0)}$ is the total mass.
Expressions (\ref{eq:c3}) and (\ref{eq:c5}) are undefined in $\mathcal{D}=2$, to be replaced by
\begin{equation}\label{eq:c6} 
 \mathcal{U}(r_2)-\mathcal{U}(r_1)=G\big(\pi\rho_\mathrm{m}^{(0)}r_1^2\big)\ln\frac{r_2}{r_1}
 \quad: \mathcal{D}=2,
\end{equation}
\begin{equation}\label{eq:c7} 
U_\mathrm{S}=\frac{2Gm_\mathrm{t}^2}{r_\mathrm{s}^4}\int_0^\infty
dr_2\,\rho(r_2)r_1^2r_2\ln\frac{r_2}{r_1}
\quad: \mathcal{D}=2.
\end{equation}
Using reference values introduced previously we arrive at the following expressions for the scaled self-energy,
$\hat{U}_\mathrm{S}\doteq U_\mathrm{S}/Np_\mathrm{s}V_\mathrm{c}$:
\begin{equation}\label{eq:c8} 
\hat{U}_\mathrm{S}=\left\{
\begin{array}{l}
{\displaystyle \frac{2\mathcal{D}}{\mathcal{D}-2}\int_0^\infty 
d\hat{r}_2\,\rho(\hat{r}_2)
\Big[\hat{r}_1^2\hat{r}_2^{\mathcal{D}-1}-\hat{r}_1^\mathcal{D}\hat{r}_2\Big]}, 
\rule[-2mm]{0mm}{10mm} \\ \rule[-2mm]{0mm}{10mm} 
{\displaystyle 4\int_0^\infty
d\hat{r}_2\,\rho(\hat{r}_2)\hat{r}_1^2\hat{r}_2\ln\frac{\hat{r}_2}{\hat{r}_1}}
\quad: \mathcal{D}=2.
\end{array} \right.
\end{equation}

In $\mathcal{D}=3$ the change $\Delta\hat{U}_\mathrm{S}$ between any two macrostates must be identical to the commonly used expression $\Delta U_\mathrm{S}^{(\mathrm{F})}$ constructed as described earlier.
If one of the macrostates is the ILG ground state we have
\begin{equation}\label{eq:c9} 
\Delta U_\mathrm{S}=\frac{3Gm_\mathrm{t}^2}{r_\mathrm{s}^{6}}
\int_0^\infty dr_2\,\rho(r_2)
\Big[r_1^2r_2^2-r_1^3r_2\Big]
\end{equation}
and
\begin{equation}\label{eq:c10} 
\Delta U_\mathrm{S}^{(\mathrm{F})}=\frac{1}{2}
\int_0^\infty dr(4\pi r^2)\big[\rho_\mathrm{m}(r)\mathcal{U}_\mathrm{F}(r)
-\rho_\mathrm{m}^{(0)}(r)\mathcal{U}_\mathrm{F}^{(0)}(r)\big], 
\end{equation}
where 
\begin{equation}\label{eq:c11}
\mathcal{U}_\mathrm{F}(r)=-\int_\infty^rdr'g(r'),\quad g(r)=-G\frac{m_\mathrm{in}(r)}{r^2},
\end{equation}
\begin{equation}\label{eq:c12}
 m_\mathrm{in}(r)=\int_0^r dr'\big(4\pi r'^2\big)\rho_\mathrm{m}(r'),
\end{equation}
and analogous expressions for the ground-state mass density, $\rho_\mathrm{m}^{(0)}(r)=\rho_\mathrm{m}^{(0)}\theta(r_\mathrm{s}-r)$.
For a symmetric cluster in $\mathcal{D}=3$ with mass confined to radius $R$, the potential $\mathcal{U}(r)$ used in (\ref{eq:c1}) and the potential $\mathcal{U}_\mathrm{F}(r)$ used in (\ref{eq:c10}) are related by a mere shift as follows \cite{note4}:
\begin{equation}\label{eq:c17} 
\mathcal{U}_\mathrm{F}(r)=\mathcal{U}(r)
-\frac{d}{d r}\Big[r\mathcal{U}(r)\Big]_{r=R}.
\end{equation}

One formal proof of $\Delta U_\mathrm{S}=\Delta U_\mathrm{S}^{(\mathrm{F})}$ in $\mathcal{D}=3$ proceeds as follows.
We begin by bringing (\ref{eq:c9}) into a form that is a better target for (\ref{eq:c10}):
\begin{align}\label{eq:c13} 
\Delta U_\mathrm{S} &=\frac{\alpha}{3}\int_0^\infty dr\rho(r)
\big[r_1^2(r)r^2-r_1^3(r)r\big] \nonumber \\
&=\frac{\alpha}{15}\,r_\mathrm{s}^5-\alpha\int_0^\infty dr\,r\rho(r)
\int_0^r dr'\,r'^2\rho(r'),
\end{align}
where we have defined $\alpha\doteq(4\pi\rho_\mathrm{m}^{(0)})^2G$, used (\ref{eq:c4}) in the second term of the square bracket, and used the derivative of (\ref{eq:c4}), $dr_1/dr=\rho(r)r^2/r_1^2$, in the first term.

Next we split up the two terms of (\ref{eq:c10}) as $\Delta U_\mathrm{S}^{(\mathrm{F})}=\Delta U_1^{(\mathrm{F})}-\Delta U_0^{(\mathrm{F})}$ with
\begin{equation}\label{eq:c14} 
\Delta U_1^{(\mathrm{F})}=\frac{\alpha}{2}\int_0^\infty \!\!dr\,r^2\rho(r)
\int_\infty^r\frac{dr'}{r'^2}\int_0^{r'} \!\!dr''\,r''^2\rho(r'')
\end{equation}
and the same expression for $\Delta U_0^{(\mathrm{F})}$  but with $\rho(r)$ and $\rho(\bar{r})$ replaced by $\theta(r_\mathrm{s}-r)$ and $\theta(r_\mathrm{s}-r'')$, respectively.
Integrating (\ref{eq:c14}) by parts yields the expression,
\begin{align}\label{eq:c15} 
 \Delta U_1^{(\mathrm{F})} 
 &=\frac{\alpha}{2}\int_0^\infty dr\,r^2\rho(r) \\
 &\hspace{5mm}\times\left[\int_\infty^r dr'r'\rho(r')
 -\frac{1}{r}\int_0^r dr'r'^2\rho(r')\right], \nonumber 
\end{align}
of which the second term is equal to half the integral term in (\ref{eq:c13}).
The other half comes from the first term in (\ref{eq:c15}), as becomes evident after interchanging the sequence of integration in the sector of $r,r'$.
That leaves the (elementary) double integral of $\Delta U_0^{(\mathrm{F})}$, which matches the first term in (\ref{eq:c13}).

%
\section{Solid-gas approximation}\label{sec:appd}
%
At very low temperatures, the numerical solution of (\ref{eq:63}) yields density profiles that include a narrow interface between a solid-like core surrounded by a dilute gas.
Replacing that interface by a solid-gas phase boundary greatly simplifies the analysis and predicts density profiles that connect with the results derived from (\ref{eq:63}) at very low $\hat{T}$ \cite{CS98, PW05}.

Technically, we substitute,  in (\ref{eq:56}), $m_\mathrm{in}$ from (\ref{eq:119}) with $m_\mathrm{t}$ (total mass), assuming that the gas contributes negligibly.
We also replace the ILG EOS (\ref{eq:120})  by its ICG limit.
With these ingredients (\ref{eq:121}) can then be solved by separation of variables
\begin{equation}\label{eq:84} 
\int_{p_\mathrm{i}}^{p}\frac{dp'}{p'}= 
-\frac{Gm_\mathrm{t}m_\mathrm{c}}{k_\mathrm{B}T}\int_{r_\mathrm{i}}^{r}dr'\,r'^{1-\mathcal{D}},
\end{equation}
where $p_\mathrm{i}$ and $\rho_\mathrm{i}=p_\mathrm{i}V_\mathrm{c}/k_\mathrm{B}T$ are the pressure and the density of the gas, respectively, at the interface located at radius $r_\mathrm{i}$. 

In $\mathcal{D}=1$ this solid-gas approximation confirms the exponential decay profile (\ref{eq:85}) throughout the gas phase.
The solution of (\ref{eq:84}) in $\mathcal{D}=2$ also confirms the power-law asymptotics (\ref{eq:82}) but supplies no hint of a critical temperature.
In $\mathcal{D}=3$ the leveling-off asymptotics, $\rho\sim e^{2/\hat{r}\hat{T}}$, predicted by (\ref{eq:84}), is consistent with the need of a wall-confinement to stabilize clusters of finite mass. 

%
\section{Decay laws}\label{sec:appa}
%
The asymptotic decay of the density profile at large distances from the center of a finite or infinite cluster is amenable to exact analysis.
Stable clusters of finite mass at $\hat{T}>0$ only exist in dimensions $\mathcal{D}\leq2$, the condition being that $\lim_{\hat{r}\to\infty}\hat{\mathcal{U}}(\hat{r})=\infty$.
In order to determine the decay law of $\rho(\hat{r})$ at $\hat{r}\to\infty$ we then convert (\ref{eq:125}) into
\begin{equation}\label{eq:a1} 
\frac{\hat{r}^{\mathcal{D}-1}\rho'(\hat{r})}{\rho(\hat{r})[1-\rho(\hat{r})]}
=-\frac{2}{\hat{T}}
\left[1-\mathcal{D}\int_{\hat{r}}^\infty d\hat{r}'\hat{r}'^{\mathcal{D}-1}\rho(\hat{r}')\right],
\end{equation}
where we have used (\ref{eq:67}).
If the decay rate is of the type $\rho\sim\hat{r}^{-(\mathcal{D}+\epsilon)}$ with $\epsilon>0$ or faster we can infer from (\ref{eq:a1}) the relation
\begin{equation}\label{eq:a2} 
\lim_{\hat{r}\to\infty}\hat{r}^{\mathcal{D}-1}\frac{\rho'(\hat{r})}{\rho(\hat{r})}=-\frac{2}{\hat{T}}.
\end{equation}

In $\mathcal{D}=1$, where a stable cluster exists at all finite $\hat{T}$, the solution of (\ref{eq:a2}) yields the exponential decay law,
\begin{equation}\label{eq:a3} 
\rho(\hat{r})_\mathrm{as}\sim e^{-2\hat{r}/\hat{T}}\quad: \mathcal{D}=1,
\end{equation}
which confirms all evidence compiled in Sec.~\ref{sec:Deq1}.

In $\mathcal{D}=2$, stable ILG clusters exist at $\hat{T}<\hat{T}_\mathrm{c}=\frac{1}{2}$ as shown in Sec.~\ref{sec:Deq2}.
The exact decay law inferred from (\ref{eq:a2}) is the power law,
\begin{equation}\label{eq:a4} 
\rho(\hat{r})_\mathrm{as}\sim \hat{r}^{-2/\hat{T}}\quad: \mathcal{D}=2,
\end{equation}
as anticipated from numerical data and analytic results for limiting cases.

Open ILG clusters in unrestricted space exist in $\mathcal{D}=1,2,3$.
In Sec.~\ref{sec:open syst} we consider finite clusters in $\mathcal{D}=1,2$ and infinite clusters in $\mathcal{D}=3$.
For finite clusters we examine the asymptotic decay of the density profile starting again from (\ref{eq:125}), which we convert, using the scaled variables (\ref{eq:134}), into
\begin{equation}\label{eq:a5} 
\frac{\tilde{r}^{\mathcal{D}-1}\tilde{\rho}'(\tilde{r})}{\tilde{\rho}(\tilde{r})[1-\rho_0\tilde{\rho}(\tilde{r})]}
=-\nu_\mathcal{D}(\rho_0)
+\int_{\tilde{r}}^\infty d\tilde{r}'\tilde{r}'^{\mathcal{D}-1}\tilde{\rho}(\tilde{r}'),
\end{equation}
where
\begin{equation}\label{eq:a6}
\nu_\mathcal{D}(\rho_0)=\int_0^\infty d\tilde{r}'\tilde{r}'^{\mathcal{D}-1}\tilde{\rho}(\tilde{r}').
\end{equation}
For decay rates that are as rapid as suggested by the data in Figs.~\ref{fig:fig20}(a) and 9(b) we infer that
\begin{equation}\label{eq:a7}
\lim_{\tilde{r}\to\infty} \tilde{r}^{\mathcal{D}-1}\frac{\tilde{\rho}'(\tilde{r})}{\tilde{\rho}(\tilde{r})}
=-\nu_\mathcal{D}(\rho_0)
\end{equation}
holds for $\mathcal{D}=1,2$.
We proceed as earlier and obtain the decay laws
\begin{equation}\label{eq:a8}
\tilde{\rho}(\tilde{r})_\mathrm{as} \sim \left\{
\begin{array}{ll}  e^{-\nu_1(\rho_0)\tilde{r}}\quad &: \mathcal{D}=1,\rule[-2mm]{0mm}{6mm} \\
\tilde{r}^{-\nu_2(\rho_0)}\quad &: \mathcal{D}=2. \rule[-2mm]{0mm}{6mm}\end{array} \right.
\end{equation}
The quantities $\nu_\mathcal{D}(\rho_0)$ that govern the exponents in (\ref{eq:a8}) are known in the ICG limit:
\begin{equation}\label{eq:a9}
\lim_{\rho_0\to0}\nu_1(\rho_0)=\sqrt{2},\quad \lim_{\rho_0\to0}\nu_2(\rho_0)=4.
\end{equation}
Their dependence on $\rho_0$ must be determined empirically from solutions of (\ref{eq:139}) via (\ref{eq:a6}).
The results (\ref{eq:a8}) are, of course, transcribed versions of (\ref{eq:a3}) and (\ref{eq:a4}) for convenient use in Sec.~\ref{sec:open syst}.

The asymptotic decay (\ref{eq:143}) for an infinite cluster in $\mathcal{D}=3$ is more readily determined by substituting a simple power-law ansatz into (\ref{eq:139}) \cite{Chav02a}.
The leading term,
\begin{equation}\label{eq:a10}
 \tilde{\rho}(\tilde{r})_\mathrm{as}\sim 2\,\tilde{r}^{-2}\quad:~ \mathcal{D}=3,
\end{equation}
is independent of $\rho_0$.

%
\section{Exact profile in $\mathcal{D}=1$}\label{sec:appb}
%
The ODE (\ref{eq:139}) for an open ILG cluster in $\mathcal{D}=1$ is amenable to exact analysis.
We write
\begin{equation}\label{eq:b1} 
\frac{\rho''}{\rho}-\frac{1-2\rho}{1-\rho}\left(\frac{\rho'}{\rho}\right)^2
+\frac{1}{\rho_0}\rho(1-\rho)=0
\end{equation}
with $\rho(\tilde{r})=\rho_0\tilde{\rho}(\tilde{r})$ and the scaled radius $\tilde{r}$ from (\ref{eq:134}).
The boundary conditions are $\rho(0)=\rho_0$ and $\rho'(0)=0$.

In (\ref{eq:b1}) the variable $\tilde{r}$ does not appear explicitly and we know from (\ref{eq:125}) that $\rho(\tilde{r})$ must be a monotonically decreasing function.
Hence there exists a unique inverse function $\tilde{r}(\rho)$, which solves the ODE,
\begin{equation}\label{eq:b2} 
\tilde{r}''+\frac{1-2\rho}{\rho(1-\rho)}\tilde{r}'-\frac{1}{\rho_0}\rho^2(1-\rho)(\tilde{r}')^3=0,
\end{equation}
with boundary conditions $\tilde{r}(\rho_0)=0$ and $\tilde{r}'(\rho_0)=-\infty$.
The solution of this first-order ODE for $\tilde{r}'(\rho)$,
\begin{equation}\label{eq:b3} 
\tilde{r}'(\rho)=-\sqrt{\frac{\rho_0}{2}}\frac{1}{\rho(1-\rho)}\frac{1}{\sqrt{\ln\frac{1-\rho}{1-\rho_0}}},
\end{equation}
features an integrable divergence at $\rho\to\rho_0$ (third factor) and a nonintegrable divergence at $\rho\to0$ (second factor).
In the inverted profile,
\begin{equation}\label{eq:b4} 
\tilde{r}(\rho)=\sqrt{\frac{\rho_0}{2}}\int_{\rho}^{\rho_0}\frac{d\rho_1}{\rho_1(1-\rho_1)\sqrt{\ln\frac{1-\rho_1}{1-\rho_0}}}~ 
: 0\leq\rho\leq\rho_0,
\end{equation}
these divergences in $\tilde{r}'(\rho)$ account for the cusp at $\rho\to\rho_0$ and the divergence at $\rho\to0$, respectively, in $\tilde{r}(\rho)$.

The exact result (\ref{eq:b4}) can be rendered as a lengthy expression involving multilogarithmic functions.
It is readily transcribed into the solution $\hat{r}(\rho)$ of (\ref{eq:63}) by substituting $\hat{T}/2$ for $\rho_0$ in the factor before the integral.

The explicit ICG result (\ref{eq:140}) is recovered by considering the scaled variable $\tilde{\rho}=\rho/\rho_0$ in (\ref{eq:b4}) and taking the limit
\begin{equation}\label{eq:b5} 
\lim_{\rho_0\to0}\tilde{r}(\rho)=\sqrt{\frac{1}{2}}\int_{\tilde{\rho}}^1\frac{d\tilde{\rho}_1}{\tilde{\rho}_1\sqrt{1-\tilde{\rho}_1}}
=\sqrt{2}\,\mathrm{artanh}\sqrt{1-\tilde{\rho}}.
\end{equation}

The exact exponent $\nu_1(\rho_0)$ of the exponential decay law (\ref{eq:a8}) rigorously established in Appendix~\ref{sec:appa} can now be determined from (\ref{eq:b4}) via integration.
We convert the integral (\ref{eq:a6}) into
\begin{equation}\label{eq:b6} 
\nu_1(\rho_0)=\frac{1}{\rho_0}\int_0^{\rho_0}d\rho_2\tilde{r}(\rho_2).
\end{equation}
Interchanging the sequence in the double integration leads to the analytic result,
\begin{equation}\label{eq:b7} 
\nu_1(\rho_0)=\sqrt{-\frac{2}{\rho_0}\ln(1-\rho_0)},
\end{equation}
which includes the ICG limit (\ref{eq:a9}) and perfectly matches the asymptotics of the data such as used in Fig.~\ref{fig:fig20}.

By the same method, we can determine the exact density profile of the hard-rod system investigated by Champion and Alastuey \cite{CA15}, i.e. the solution of (\ref{eq:e4}) in $\mathcal{D}=1$.
The result,
\begin{align}\label{eq:b8}
  \hat{r}(\rho) & = \sqrt{\frac{\hat{T}(1-\rho_{0})}{4\rho_{0}}}
  \int_{\rho}^{\rho_{0}}
  \frac{d \rho}{\rho(1-\rho)^{2}
    \sqrt{1-\frac{(1-\rho_{0})\rho}{(1-\rho)\rho_{0}}}} 
\nonumber   \\ & = \sqrt{\frac{\hat{T}\rho_{0}}{1-\rho_{0}}}
  \sqrt{1-\frac{(1-\rho_{0})\rho}{(1-\rho)\rho_{0}}} \nonumber \\
  & +\sqrt{\frac{\hat{T}(1-\rho_{0})}{\rho_{0}}}\mathrm{artanh}
  \left(\sqrt{1-\frac{(1-\rho_{0})\rho}{(1-\rho)\rho_{0}}}\right),
\end{align}
is equivalent to expression (78) in Ref.~\cite{CA15}.

%
\section{Short-distance regularizations}\label{sec:appe}
%
Antidotes against collapse in self-gravitating systems of massive particles come in two types.
One type softens the law of gravity at short distances and thus remove forces of divergent strength \cite{DdV07, IK03, MARF17, FL00}.
The other type keeps the particles away from the divergences by a short-distance repulsive force of some kind \cite{AH72, SKS95, CA15, MARF17}.
In the following, we compare consequences of short-distance regularization for three realizations of the latter type.

The first realization is, of course, the ILG with EOS (\ref{eq:60}), which produces ODE (\ref{eq:63}) for the density profile, both reproduced here for easy reference:
\begin{equation}\label{eq:e1} 
\hat{p}(\hat{r})=-\hat{T}\ln\big(1-\rho(\hat{r})\big),
\end{equation}
\begin{equation}\label{eq:e2} 
\frac{\rho''}{\rho} +\frac{\mathcal{D}-1}{\hat{r}}\frac{\rho'}{\rho} -\frac{1-2\rho}{1-\rho}\left(\frac{\rho'}{\rho}\right)^2 
+\frac{2\mathcal{D}}{\hat{T}}\rho(1-\rho)=0.
\end{equation}

The second realization uses the EOS,
\begin{equation}\label{eq:e3} 
\hat{p}(\hat{r})=\frac{\hat{T}\rho(\hat{r})}{1-\rho(\hat{r})},
\end{equation}
which is more strongly divergent than (\ref{eq:e1}) as the limit of maximum density is approached.
The EOS (\ref{eq:e3}) is exact in $\mathcal{D}=1$ for hard rods in a continuum as shown by Champion and Alastuey \cite{CA15} (see also Ref.~\cite{inharo}) and is commonly used in $\mathcal{D}=3$ as a phenomenological EOS for hard-sphere models \cite{AH72, SKS95}.
When we carry out the analysis of Sec.~\ref{sec:dif-equ} with this EOS we arrive at the ODE,
 \begin{equation}\label{eq:e4} 
\frac{\rho''}{\rho} +\frac{\mathcal{D}-1}{\hat{r}}\frac{\rho'}{\rho} -\frac{1-3\rho}{1-\rho}\left(\frac{\rho'}{\rho}\right)^2 
+\frac{2\mathcal{D}}{\hat{T}}\rho(1-\rho)^2=0,
\end{equation}
which differs from (\ref{eq:e2}) in the last two terms.
The scaling for this case is explained in Ref.~\cite{inharo}.

The third realization is the FD gas, where an effective short-distance repulsion is produced by the Pauli exclusion principle.
The EOS in parametric form reads,
\begin{equation}\label{eq:e5} 
p=\frac{g_sk_\mathrm{B}T}{\lambda_T^\mathcal{D}}f_{\mathcal{D}/2+1}(z),\quad
\rho=g_sf_{\mathcal{D}/2}(z),
\end{equation}
where $g_s$ is the spin degeneracy, $z$ the fugacity, $\lambda_T\doteq(h^2/2\pi m_\mathrm{c}k_\mathrm{B}T)^{1/2}$ the thermal wavelength, and 
\begin{equation}\label{eq:e6} 
f_n(z)\doteq\frac{1}{\Gamma(n)}\int_0^\infty dx\frac{x^{n-1}}{z^{-1}e^x+1},
\end{equation}
the Fermi-Dirac function.
In this case, the analysis leads to the ODE \cite{note5}
 \begin{equation}\label{eq:e7} 
\frac{z''}{z} +\frac{\mathcal{D}-1}{\hat{r}}\frac{z'}{z} -\left(\frac{z'}{z}\right)^2 
+\frac{2\mathcal{D}}{\hat{T}}f_{\mathcal{D}/2}(z)=0,
\end{equation}
for the fugacity profile $z(\hat{r})$, from which the density profile follows via $\rho(\hat{r})=g_sf_{\mathcal{D}/2}\big(z(\hat{r})\big)$.
The scaling for this case replaces $V_\mathrm{c}$ by $\lambda_T^\mathcal{D}$ in (\ref{eq:2})-(\ref{eq:3}).

In the low-density regime, $\rho\ll1$, all three ODEs reduce to the ODE (\ref{eq:68}) for the ICG, which is not surprising but instructive nevertheless.
One conclusion is that the asymptotic decay laws (\ref{eq:146}) in $\mathcal{D}=1,2,3$ hold for all three realizations and are, in all likelihood, universal, i.e. independent of how the short-distance regularization is implemented. 

A second conclusion is that the continuous phase transition in $\mathcal{D}=2$ described for the ILG in Sec.~\ref{sec:Deq2} is also exact for all three cases and is very likely to be universal in the same sense too.
That transition invariably takes place in the low-density limit, where the density profile is governed by the ICG ODE (\ref{eq:68}).
The scaled temperature is invariant, only the reference temperatures are model-dependent.

The discontinuous phase transition in $\mathcal{D}=3$ described for the ILG in Sec.~\ref{sec:Deq3}, by contrast, takes place, in general, away from the ICG limit. 
It is, therefore, expected to exhibit features that depend on the specific short-distance regularization. 
The differences noted between ILG and ICG regarding this transition illustrate this point.

Also in $\mathcal{D}=3$, the ODEs for all three realizations of short-distance regularization reproduce the outermost layer, the halo, of the (iconic) Bonnor-Ebert sphere with its characteristic $\sim \hat{r}^{-2}$ decay in the density profile.
However, every realization produces its distinctive structure closer to the center of the (infinite) cluster. 
Qualitative similarities between the ILG and the FD gas have been noted in Sec.~\ref{sec:Deq3op}.

Finally, for $\mathcal{D}=1$, we have calculated the exact density profiles (\ref{eq:b4}) for the ILG realization and (\ref{eq:b8}) for the hard-rod realization analyzed in Ref.~\cite{CA15}, namely the solutions of (\ref{eq:e2}) and (\ref{eq:e4}), respectively. 
These profiles differ in shape at $T>0$ on account of the differences in (\ref{eq:e2}) and (\ref{eq:e4}).
However, both exhibit exponential decay on account of their common ICG limit.
The corresponding profile of the FD-gas realization, i.e. the function $\rho(\hat{r})$ inferred from the solution of (\ref{eq:e7}), is known to have a nontrivial shape even at $T=0$, yet its exponential asymptotics at $T>0$ is the same again. 




\begin{thebibliography}{100}
\bibitem{note4}
For the derivation of (\ref{eq:c17}) we have used (\ref{eq:88}), the known $\mathcal{U}_\mathrm{F}(r)=Gm_\mathrm{t}/r$ at $r>R$, and the requirements that both 
$\mathcal{U}(r)$ and $\mathcal{U}_\mathrm{F}(r)$ are continuous and smooth at $r=R$.
\bibitem{note5}
A note of caution: (\ref{eq:e7}) only holds for clusters with planar symmetry in $\mathcal{D}=1$, clusters with cylindrical symmetry in $\mathcal{D}=2$, and clusters with spherical symmetry in $\mathcal{D}=3$. Unlike (\ref{eq:e2}) for the ILG, it does not hold for clusters with with planar or cylindrical symmetry in $\mathcal{D}=3$ or for clusters with planar symmetry in $\mathcal{D}=2$.
\end{thebibliography}
\end{document}